\documentclass[10pt,dvips,a4paper]{article}
\usepackage[pdftex]{graphicx}
\include{commands}

\usepackage{amssymb}
\usepackage{amsmath}
\usepackage{setspace}
\usepackage[round]{natbib}
\usepackage[pdftex]{graphicx,xcolor}

\renewcommand{\title}{Evaluation of nonlocal approaches for modelling fracture near nonconvex boundaries}

\begin{document}

\begin{center} \begin{LARGE} \textbf{\title} \end{LARGE} \end{center}

\begin{center} Peter Grassl$^{\mbox{1}}$, Dimitris Xenos$^{\mbox{1}}$, Milan Jir\'{a}sek$^{\mbox{2}}$ and Martin Hor\'{a}k$^{\mbox{2}}$  \\

$^{1}$School of Engineering, University of Glasgow, Glasgow, UK\\

$^{2}$Department of Mechanics, Faculty of Civil Engineering, Czech Technical University in Prague, Czech Republic

\today

\end{center}

\section*{Abstract}
Integral-type nonlocal damage models describe the fracture process zones by regular strain profiles insensitive to the size of finite elements, which is achieved by incorporating weighted spatial averages of certain state variables into the stress-strain equations. 
However, there is no consensus yet how the influence of boundaries should be taken into account by the averaging procedures.
In the present study, nonlocal damage models with different averaging procedures are applied to the modelling of fracture in specimens with various boundary types.
Firstly, the nonlocal models are calibrated by fitting load-displacement curves and dissipated energy profiles for direct tension to the results of mesoscale analyses performed using a discrete model.
These analyses are set up so that the results are independent of boundaries.
Then, the models are applied to two-dimensional simulations of three-point bending tests with a sharp notch, a  V-type notch, and a smooth boundary without a notch.
The performance of the nonlocal approaches in modelling of fracture near nonconvex boundaries is evaluated by comparison of load-displacement curves and dissipated energy profiles along the beam ligament with the results of meso-scale simulations. As an alternative approach,
elastoplasticity combined with nonlocal or over-nonlocal damage is also included in the comparative study.

\section{Introduction}
Among other important factors, the failure process of concrete strongly depends on the meso-structure. 
Growth and coalescence of microcracks lead to the formation of fracture process zones (FPZ) which transfer stresses by crack bridging and aggregate interlock. 
Inelastic processes in such zones are commonly modelled by nonlinear fracture mechanics, e.g.\ cohesive crack models, or by continuum damage mechanics using stress-strain laws with strain softening.

One group of continuum damage mechanics approaches suitable for computational structural analysis are integral-type nonlocal models, which describe the localised fracture process zones by regular strain profiles independently of the size of finite elements \citep{PijBaz87,BazJir02}.
This is achieved by evaluating the stress at each point based on weighted averages of state variables in the vicinity of that point.
However, there is no consensus on how the averaging should be adjusted near the physical boundary of the body. 
Commonly used scaling procedures may result in excessive spurious energy dissipation close to boundaries for notched specimens \citep{JirRolGra04}. 
In this previous study, it was suggested that the excess in dissipated energy originates from including the contribution of the undamaged material below the notch to the nonlocal variable at a point above the notch, which reduces the damage and, therewith, introduces an artificial strengthening at this point. 
In alternative approaches which have the potential to reduce this spurious effect, the averaging procedure depends on the distance to boundaries \citep{BolHik95,KraPijDuf09,BazLeHo10}, or on the stress state  \citep{Baz94,JirBaz94,GirDufMaz11}. 
Another formulation, which preserves symmetry of the nonlocal weight function, was proposed by \cite{Pol02},  \cite{BorFaiPar02} and \cite{BorFaiPar03} and will be called here the method of local complement.
In addition, the spurious energy dissipation might also be affected by the choice of more advanced constitutive models, such as elasto-plasticity combined with nonlocal damage \citep{GraJir06a,Gra09b}, where the plastic part could be expected to limit the effective stress and therewith reduce the artificial strengthening described above.

In the present work, a nonlocal damage model with four averaging procedures (representing standard, distance-based, stress-based and local complement averaging) and a plasticity model with nonlocal damage based on standard and over-nonlocal averaging procedures \citep{VerBri94,Stromberg96,GraJir06} are applied to the modelling of fracture in notched concrete beams subjected to three-point bending.
Initially, the models are calibrated by fitting meso-scale analysis results obtained for a problem independent of boundaries \citep{GraJir10}. 
Thus, this calibration is unaffected by the type of averaging procedure used. 
Then, the nonlocal models are applied to two-dimensional simulations of three-point bending beams with a sharp notch, a V-type notch and a smooth boundary without a notch, for which the averaging procedures close to boundaries are expected to influence the response. 
The results of the different models are presented in terms of load-displacement curves and dissipated energy profiles, and again compared with meso-scale analyses results.

The meso-scale analyses, which are used to produce reference results to compare the nonlocal model results with, are based on mapping the material properties of individual phases of the heterogenous meso-structure of concrete on a background mesh \citep{SchMie92b}.
The fracture process of the background mesh is described as the progressive failure of discrete elements, such as lattices of bars and beams \citep{Kaw78,Cun79}.
In lattice approaches, the connectivity between nodes is not changed so that contact determination is simplified.
Lattice models are mainly suitable for analyses involving small strains \citep{HerHanRou89,SchMie92b,BolSai98}.
In recent years, the discrete element method based on a lattice determined by the Voronoi tessellation has been shown to be suitable for modelling fracture \citep{BolSai98}.
The constitutive response for the individual phases can be described by micro-mechanics or phenomenological constitutive models, commonly based on the theory of plasticity, damage mechanics, or a combination of the two.
For predominantly tensile loading, an isotropic damage model has shown to provide satisfactory results \citep{GraJir10}. Such a model is used here for the meso-scale simulations.

The present meso-scale analyses are based on several assumptions.
In the chosen idealisation of the meso-structure only large aggregates are considered, and are embedded in a mortar matrix separated by interfacial transition zones.
The aggregates are assumed to be linear elastic and stiffer than the matrix, whereas the interfacial transition zone is assumed to be weaker and more brittle than the matrix.
The material constants for the constitutive models of the three phases are chosen by comparing the global results of analyses and experiments assuming certain ratios of the properties of different phases.
For instance, aggregates are assumed to be twice as stiff as the matrix, which in turn is twice as strong and ductile as the interfacial transition zone. 
These chosen ratios are supported by experimental results reported in the literature \citep{HsuSla63} and were used in a recent study on the size effect in notched concrete beams in \cite{GraGreSol12} for which the analysis results were in good agreement with experimental data.
Furthermore, the present study is limited to two-dimensional plane stress analyses with aggregates idealised as circular inclusions. 
These are of course strong simplifications. Nevertheless, it is believed that even such an idealised meso-scale model reflects the main features of the mechanical behaviour of concrete as a heterogeneous material with stiff inclusions in a quasi-brittle matrix. 
In the absence of detailed experimental measurements of the effect of boundaries on the process zone size and energy dissipation density,  the meso-scale model is used in the present study as a reference solution against which the nonlocal models are compared.

The meso-scale model reflects the interactions that take place at the material scale and the resulting local redistributions of stress and strain fluctuations. In the nonlocal continuum model, such effects are taken into account in an approximate and simplified way by weighted spatial averaging of an internal variable linked to the inelastic processes.

\section{Macroscopic models}\label{sec:macroscopicmodel}
In the present section, two macroscopic nonlocal constitutive models based on damage mechanics and on a combination of plasticity and damage mechanics are briefly summarised in sections~\ref{sec:damage}~and~\ref{sec:plasticdamage}, respectively.
Then, the different averaging procedures are described in section~\ref{sec:boundaries}.

\subsection{Damage model} \label{sec:damage}
The total stress-strain relationship for the isotropic damage model is
\vskip-.6cm
\begin{eqnarray}\label{eq:stressStrainDamage}
\boldsymbol{\sigma} &=& (1-\omega) \mathbf{D}_{\rm e}: \boldsymbol{\varepsilon}= (1-\omega) \tilde{\boldsymbol{\sigma}}
\end{eqnarray}
where $\boldsymbol{\sigma}$ is the total stress tensor, $\omega$ is the damage variable, $\mathbf{D}_{\rm e}$ is the isotropic elastic stiffness tensor based on Young's modulus $E$ and Poisson's ratio $\nu$, $\boldsymbol{\varepsilon}$ is the strain and $\tilde{\boldsymbol{\sigma}}$ is the effective stress tensor.
Damage is driven by a history variable $\kappa_{\rm d}$ and is determined by
the damage law
\vskip-.6cm
\begin{eqnarray}\label{eq:damageDamage}
\omega\left(\kappa_{\rm d}\right) =\left\{
\begin{array}{ll}
1-\exp\left(-\displaystyle\frac{1}{m_{\rm d}} \left(\frac{\kappa_{\rm d}}{\varepsilon_{\rm{max}}}\right)^{m_{\rm d}}\right)&\mbox{, $\kappa_{\rm d} \leq \varepsilon_1$}\\[3mm]
1- \displaystyle\frac{\varepsilon_3}{\kappa_{\rm d}}\exp\left(-\frac{\kappa_{\rm d}-\varepsilon_1}{\varepsilon_{\rm f}\left[1+\left(\frac{\kappa_{\rm d}-\varepsilon_1}{\varepsilon_2}\right)^n\right]}\right)&\mbox{, $\kappa_{\rm d} > \varepsilon_1$}\\
\end{array}
\right.
\end{eqnarray}
where
\vskip-.6cm
\begin{eqnarray}\label{eq:md}
m_{\rm d} &=& \dfrac{1}{\ln(E \varepsilon_{\rm max}/f_{\rm t})}
\end{eqnarray}
and $f_{\rm t}$ is the uniaxial tensile strength.
Parameter $\varepsilon_{\rm max}$ is the axial strain at peak stress, and $\varepsilon_1$, $\varepsilon_2$ and $n$ are additional parameters that control the softening part of the stress-strain diagram.
Furthermore,
\vskip-.6cm
\begin{eqnarray}
\varepsilon_{\rm f} &=& \dfrac{\varepsilon_{\rm 1}}{(\varepsilon_1/\varepsilon_{\rm{max}})^{m_{\rm d}} -1}
\end{eqnarray}
and
\vskip-.6cm
\begin{eqnarray}
\varepsilon_3 &=& \varepsilon_1\exp\left(-\dfrac{1}{m_{\rm d}}\left(\dfrac{\varepsilon_1}{\varepsilon_{\rm{max}}}\right)^{m_{\rm d}}\right)
\end{eqnarray}
This damage law exhibits pre- and post-peak nonlinearities in uniaxial tension.

The history variable $\kappa_{\rm d}$, used in (\ref{eq:damageDamage}) to obtain the damage parameter, represents the maximum level of nonlocal equivalent strain $\bar{\varepsilon}_{\rm eq}$ reached in the history of the material. 
It is determined by the loading-unloading conditions 
\vskip-.6cm
\begin{eqnarray}
f \leq 0,& \dot{\kappa}_{\rm d} \geq 0,& \dot{\kappa}_{\rm d} f = 0
\end{eqnarray}
in which
\vskip-.6cm
\begin{eqnarray}
f\left(\bar{\varepsilon}_{\rm eq},\kappa_{\rm d}\right) &=& \bar{\varepsilon}_{\rm eq} - \kappa_{\rm d}
\end{eqnarray}
is the loading function.

The nonlocal equivalent strain is defined as
\vskip-.6cm
\begin{eqnarray}\label{eq:nonlocalDamage}
\bar{\varepsilon}_{\rm eq}\left(\boldsymbol{x}\right) = \int_{V} \alpha \left(\boldsymbol{x},\boldsymbol{\xi}\right) \varepsilon_{\rm eq}(\boldsymbol{\xi}) \rm{d}\boldsymbol{\xi}
\end{eqnarray} 
Here, $\boldsymbol{x}$ is the point at which the nonlocal equivalent strain $\bar{\varepsilon}_{\rm eq}$ is evaluated as a weighted average of local equivalent strains  $\varepsilon_{\rm eq}$ at all points $\boldsymbol{\xi}$ in the vicinity of $\boldsymbol{x}$ within the integration domain $V$.

According to the standard scaling approach \citep{PijBaz87},
the weight function 
\vskip-.6cm
\begin{eqnarray} \label{eq:basicWeight}
\alpha\left(\boldsymbol{x},\boldsymbol{\xi}\right) = \dfrac{\alpha_0\left(\boldsymbol{x},\boldsymbol{\xi}\right)}{\int_V \alpha_0\left(\boldsymbol{x},\boldsymbol{\xi}\right) \rm{d}\boldsymbol{\xi}}
\end{eqnarray} 
is constructed from a function  $\alpha_0(\boldsymbol{x},\boldsymbol{\xi})$ normalised by its integral over the integration domain $V$ such that the averaging scheme does not modify a uniform field.
The function 
\vskip-.6cm
\begin{eqnarray}\label{eq:basicAlpha0}
\alpha_0\left(\boldsymbol{x},\boldsymbol{\xi}\right) = \exp\left(-\dfrac{\|\boldsymbol{x} - \boldsymbol{\xi}\|}{R}\right)
\end{eqnarray}
is defined here as an exponential (Green-type) function, with parameter $R$ reflecting the 
internal material length. Modifications of the standard averaging scheme will be described in Section~\ref{sec:boundaries}.

The local equivalent strain in (\ref{eq:nonlocalDamage}) is
\vskip-.4cm
\begin{eqnarray}
\varepsilon_{\rm eq} = \dfrac{1}{E} \max_{I=1,2,3} \tilde{\sigma}_{I}
\end{eqnarray}
where $\tilde{\sigma}_{I}$ denotes the $I$-th principal component of the effective stress tensor $\tilde{\boldsymbol{\sigma}}=\mathbf{D}_{\rm e}: \boldsymbol{\varepsilon}$ introduced in (\ref{eq:stressStrainDamage}).
This constitutive law results in a Rankine strength envelope.

\subsection{Plastic-damage model} \label{sec:plasticdamage}
The second constitutive approach is based on a combination of plasticity and damage mechanics. 
The stress-strain relation for the damage-plasticity model is
\begin{equation}\label{eq:totStress}
\boldsymbol{\sigma} = (1-\omega) \tilde{\boldsymbol{\sigma}} = (1-\omega) \boldsymbol{D}_{\rm e} : \left(\boldsymbol{\varepsilon} - \boldsymbol{\varepsilon}_{\rm p}\right)
\end{equation}
where $\boldsymbol{\varepsilon}_{\rm p}$ is the plastic strain tensor.
The plasticity part is based on the effective stress tensor, and damage is driven by the plastic strain evolution.

A Rankine yield surface is used, which, in plane stress, is described by the two yield functions
\begin{equation}
f_1(\tilde{\boldsymbol{\sigma}}, \sigma_{\rm y}) = \tilde{\sigma}_{1}(\tilde{\boldsymbol{\sigma}}) - \sigma_{\rm y}
\end{equation}
\begin{equation}
f_2(\tilde{\boldsymbol{\sigma}}, \sigma_{\rm y}) = \tilde{\sigma}_{2}(\tilde{\boldsymbol{\sigma}}) - \sigma_{\rm y}
\end{equation}
Here, $\tilde{\sigma}_1$ and $\tilde{\sigma}_2$ are the principal values of the effective stress tensor $\tilde{\boldsymbol{\sigma}}$.

The yield stress $\sigma_{\rm y}$ is given by the hardening law
\begin{equation} \label{eq:yieldStress}
  \sigma_{\rm y}(\kappa_{\rm p}) = \left\{
    \begin{array}{ll}
        E_{0} \kappa_{\rm p} \exp\left(-\dfrac{1}{m_{\rm p}} \left(\dfrac{\kappa_{\rm p}}{\varepsilon_{\rm{p,max}}}\right)^{m_{\rm p}}\right) & \mbox{if  $\kappa_{\rm p} \leq \varepsilon_{\rm{p,max}}$}\\
       f_{\rm t} & \mbox{if $\kappa_{\rm p}>\varepsilon_{\rm{p,max}}$}
     \end{array}
   \right.
\end{equation}
where $\varepsilon_{\rm{p,max}}$ is the plastic strain at $f_{\rm t}$ and $\kappa_{\rm p}$ is the plastic hardening variable (cumulative plastic strain) defined by the rate equation $\dot{\kappa}_{\rm p} = \|\dot{\boldsymbol{\varepsilon}}_{\rm p}\|$.
Exponent $m_{\rm p}$ is a dependent parameter given by
\begin{equation}\label{eq:mp}
 m_{\rm p}=\dfrac{1}{\ln\left(E_0 \varepsilon_{\rm p,max}/f_{\rm t}\right)}
\end{equation}

The first part of the hardening law (\ref{eq:yieldStress})
contains an exponential term of the same form as the exponential term in the first part of the damage law (\ref{eq:damageDamage}).
However, while the damage parameter is used to relate stresses to total strains, the hardening law represents the stress-plastic strain relationship.
Consequently, parameter $E_0$ in (\ref{eq:mp}) is an additional model parameter, which is independent of Young's modulus E used in (\ref{eq:md}).
It corresponds to the initial hardening modulus and its value is typically very high, to make sure that the nonlinearity at low stress levels is negligible. 

The damage variable $\omega$ in (\ref{eq:totStress}) is determined as
\begin{equation}
  \omega = g_d(\kappa_d)= \left\{
    \begin{array}{ll}
      0 & \mbox{if $\kappa_{\rm d} \leq \varepsilon_{\rm{p,max}}$}\\
      1-d_5 \exp \left(-d_1 \left(\dfrac{\kappa_{\rm d}-\varepsilon_{\rm{p,max}}}{\varepsilon_{\rm{p,max}}}\right)^{d_2}\right) - (1-d_5) \exp \left(-d_3 \left(\dfrac{\kappa_{\rm d}-\varepsilon_{\rm{p,max}}}{\varepsilon_{\rm{p,max}}}\right)^{d_4}\right) & \mbox{if $\kappa_{\rm d}>\varepsilon_{\rm{p,max}}$}
     \end{array}
   \right.
\end{equation}
with dimensionless parameters $d_1$ to $d_5$.
It is driven by an internal variable $\kappa_{\rm d}$ which, in the local version of the model, would be considered as the cumulative plastic strain $\kappa_p$. 
In the simplest nonlocal formulation, it could be taken as the nonlocal cumulative plastic strain
\begin{equation}
\bar{\kappa}_{\rm p}\left(\mathbf{x}\right) = \int_V \alpha(\|\mathbf{x}-\boldsymbol{\xi}\|) \kappa_{\rm p}(\boldsymbol{\xi}) \rm{d}\boldsymbol{\xi}
\end{equation} 
where $\alpha$ is the weight function defined in (\ref{eq:basicWeight}).
However, for a plastic hardening law with constant yield stress after saturation of hardening, such a formulation would not provide full regularisation, as discussed e.g.\ by \cite{JirRol02}. 
Full regularisation can be achieved by the so-called over-nonlocal formulation, with 
\begin{equation}
\kappa_{\rm d} = m \bar{\kappa}_{\rm p} + (1-m) \kappa_{\rm{p}}
\end{equation}
where $m$ is an additional parameter larger than 1. This approach was introduced in the context of nonlocal plasticity by \cite{VerBri94}
and \cite{Stromberg96}, and was adapted to the damage-plastic formulation by \cite{GraJir06a}. 

\subsection{Modified averaging schemes}\label{sec:boundaries}
\begin{figure}
\centering
\begin{tabular}{cc}
\includegraphics[width=4cm]{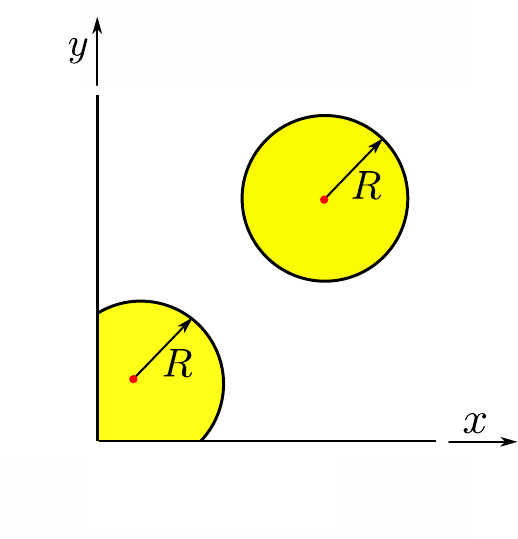} & \includegraphics[width=4cm]{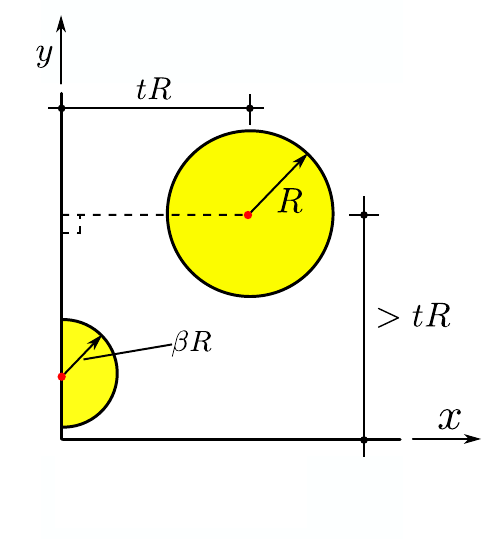}\\
(a) & (b)\\
 \includegraphics[width=4cm]{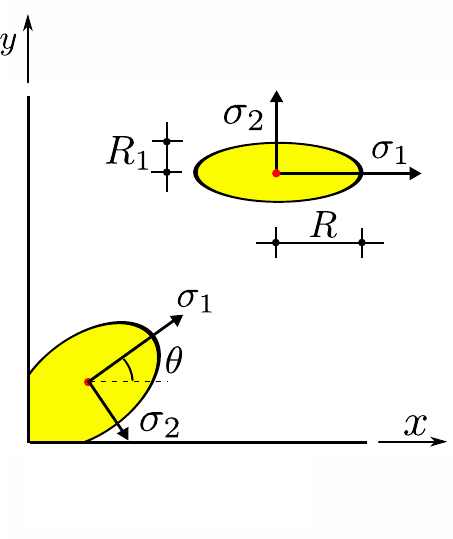} & \includegraphics[width=4cm]{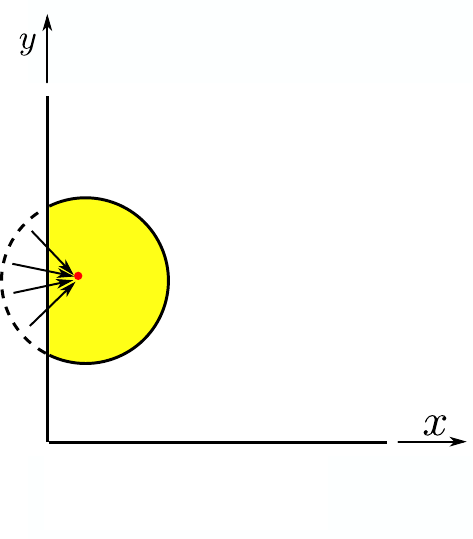}\\
(c) & (d)
\end{tabular}
\caption{Differences among four nonlocal averaging approaches near boundaries: (a) standard scaling, (b) distance-based scaling, (c) stress-based scaling, (d) local complement.}
\label{fig:boundarySchematic}
\end{figure}

This section presents three modifications of the standard averaging approach with simple rescaling according to (\ref{eq:basicWeight}).
They are referred to as the (i) distance-based approach, (ii) stress-based approach, and (iii) local complement, and they represent groups of models proposed in the literature (Fig.~\ref{fig:boundarySchematic}).
Although the distance-based and stress-based averaging approaches are based on concepts previously reported, they have not been used before in the specific formats adopted here. 

For instance, a distance-based averaging approach was already proposed by \cite{BolHik95}, who applied it to a predefined elliptic zone around a sharp notch. 
The present distance-based approach is applied to the entire specimen and, unlike \cite{KraPijDuf09}, deals with a reduced but still circular domain of influence instead of an elliptical domain. With parameter $\beta$ set to zero, the material behavior at a notch tip would be fully local while the approach of \cite{KraPijDuf09} leads to averaging along an ellipse degenerated into a straight segment.

Probably the first stress-based approach, motivated by the interaction of microcracks, was proposed by \cite{Baz94} and analyzed in a simplified form by \cite{JirBaz94}. 
In our study, the stress-based approach is phenomenological. In some aspects it is similar to the work of \cite{GirDufMaz11} who, inspired by the distance-based approach of \cite{KraPijDuf09} and by the damage-based approach of \cite{PijDuf10}, developed a more flexible concept which can handle the influence of boundaries as well as the reduction of nonlocal interactions across highly damaged zones.

The principal objective of the present paper is not to develop new models but to compare the performance of different groups of models in application to fracture close to nonconvex boundaries. The main motivation for choosing the specific details of the adopted approaches was that they are, in the authors' opinion, conceptually simple and easy to implement.  

For the {\bf standard scaling approach}, the weight function $\alpha(\boldsymbol{x},\boldsymbol{\xi})$ is scaled according to (\ref{eq:basicWeight}), with the integration domain $V$ in the denominator of (\ref{eq:basicWeight}) corresponding to the specimen under consideration.
The basic weight function  $\alpha_0(\boldsymbol{x},\boldsymbol{\xi})$ defined in
 (\ref{eq:basicAlpha0}) depends only on the distance between points 
$\boldsymbol{x} $ and $\boldsymbol{\xi}$.
Its scaling according to (\ref{eq:basicWeight}) ensures that the nonlocal operator does not alter a uniform field. 

For the {\bf distance-based approach}, the weight function is also scaled, but the basic weight function $\alpha_0(\boldsymbol{x},\boldsymbol{\xi})$  is made dependent on the minimum distance of point $\boldsymbol{x}$ to the specimen boundary (Fig.~\ref{fig:boundarySchematic}b):
\vskip-.4cm 
\begin{eqnarray}
\alpha_0\left(\boldsymbol{x},\boldsymbol{\xi}\right) &=& \exp\left(-\dfrac{\|\boldsymbol{x}-\boldsymbol{\xi}\|}{\gamma\left(\boldsymbol{x}\right) R}\right)
\end{eqnarray}
where
\vskip-.6cm 
\begin{eqnarray}\label{eq:gamma}
\gamma\left(\boldsymbol{x}\right) = \left\{ 
\begin{array}{ll}
1 & \mbox{, $d(\boldsymbol{x}) \geq t R$} \\
\dfrac{1-\beta}{t R}d(\boldsymbol{x}) +\beta & \mbox{, $d(\boldsymbol{x}) < t R$} \\
\end{array}
\right.
\end{eqnarray}
Here, $\beta$ and $t$ are parameters of the distance-based scaling approach and $d(\boldsymbol{x})$ is the minimum distance of point $\boldsymbol{x}$ to the specimen boundary. 
For a material point $\boldsymbol{x}$ lying on the boundary, the distance $d(\boldsymbol{x}) = 0$ and formula (\ref{eq:gamma}) yields $\gamma\left(\boldsymbol{x}\right) = \beta$. 
On the other hand, when the distance is greater than $t R$, $\gamma\left(\boldsymbol{x}\right) = 1$ and the present distance-based approach gives the same result as the standard scaling approach. In the boundary layer of thickness $tR$, the value of $\gamma$ varies linearly between $\beta$ and 1.

The {\bf stress-based scaling approach} (Fig.~\ref{fig:boundarySchematic}c) exploits
a transformation matrix
\vskip-.6cm     
\begin{eqnarray}\label{eq:transMatrix}
\mathbf{T} = 
\begin{pmatrix}
1 & 0\\
0 & \frac{1}{\gamma}
\end{pmatrix}
\begin{pmatrix}
n_{1x} & n_{1y}\\
-n_{1y} & n_{1x}
\end{pmatrix}
= \begin{pmatrix}
n_{1x} & n_{1y}\\
-\frac{n_{1y}}{\gamma} & \frac{n_{1x}}{\gamma}
\end{pmatrix} 
\end{eqnarray}
where $n_{1x}$ and $n_{1y}$ are the components of the unit eigenvector $\boldsymbol{n}_1$ associated with the maximum principal value of the effective stress $\tilde{\boldsymbol{\sigma}}$. Multiplication by $\mathbf{T}$ transforms
an ellipse with principal axes aligned with the principal stress directions
and principal semi-axes 1 and $\gamma$ into the unit circle.

The new function
\vskip-.6cm     
\begin{eqnarray}
\alpha_0\left(\boldsymbol{x},\boldsymbol{\xi} \right) = \exp \left(-\dfrac{ \| \mathbf{T}(\boldsymbol{x}) \cdot (\boldsymbol{\xi} - \boldsymbol{x})\| }{R}\right)
\end{eqnarray}
is affected by the effective stress at point $\boldsymbol{x}$.
In equation (\ref{eq:transMatrix}), $\gamma$ is a scaling factor, defined as
\vskip-.6cm     
\begin{eqnarray}\label{eq:stress-gamma}
\gamma = \left\{ 
\begin{array}{ll}
 \beta + (1-\beta) \left(\dfrac{\langle \tilde{\sigma}_2 \rangle}{\tilde{\sigma}_1}\right)^2 & \mbox{, $\tilde{\sigma}_1>0$} \\
1 & \mbox{, $\tilde{\sigma}_1 \leq 0$} \\
\end{array}
\right. 
\end{eqnarray}
Here, $\beta$ is a parameter of this approach, $\tilde{\sigma}_2$ is the second principal effective stress and $\langle \cdot \rangle$ denotes the MacAuley brackets (positive part operator).
For instance, for uniaxial tension the principal effective stresses are $\tilde{\sigma}_1 = \tilde{\sigma}_{\rm t}>0$ and $\tilde{\sigma}_2 =0$, which gives $\gamma = \beta$.
On the other hand, for equi-biaxial tension we have $\tilde{\sigma}_1 = \tilde{\sigma}_2 =\tilde{\sigma}_{\rm t}$, so that $\gamma = 1$, which coincides with the standard scaling approach. In this special case, $n_{1x}$ and $n_{1y}$ are not uniquely defined, but this does not matter since for $\gamma = 1$ the matrix $\mathbf{T}$ is orthogonal and $\| \mathbf{T}(\boldsymbol{x}) \cdot (\boldsymbol{\xi} - \boldsymbol{x})\|=\| \boldsymbol{\xi} - \boldsymbol{x}\|$. A minimum value of $\gamma$ is enforced by parameter $\beta$, to make sure that the contributing domain does not degenerate into a segment of zero area. Since the numerical evaluation of the nonlocal variable is still based on the Gauss integration scheme of the finite element mesh, it would be very inaccurate if the contributing domain became a too narrow ellipse.

This approach is similar to the scheme used by \cite{GirDufMaz11}, but not exactly the same. In the present approach, the modification of the nonlocal interaction weight depends on the effective stress state at the ``receiver'' point $\mathbf{x}$, whereas according to \cite{GirDufMaz11} it depends on the nominal stress state at the ``source'' point $\boldsymbol{\xi}$.

The stress-based approach does not incorporate the influence of boundaries explicitly but it can ``feel'' them through the stress field. On a free boundary of the specimen with no applied surface tractions, the principal stress directions are normal and tangential to the boundary and the principal stress perpendicular to the boundary vanishes. If the principal stress along the boundary is tensile, the internal length is reduced in the perpendicular direction. This produces a similar effect as the distance-based approach.

All the aforementioned modifications 
break the symmetry of the nonlocal weight function with respect to its
arguments  $\boldsymbol{x}$ and $\boldsymbol{\xi}$. A modification preserving
symmetry was proposed by 
\cite{Pol02},  \cite{BorFaiPar02} and \cite{BorFaiPar03},
and will be referred to as the method of {\bf local complement}.
It is based on the idea that the contribution of the ``missing'' part
of the nonlocal neighbourhood (located beyond the physical boundary of the body)
is compensated for by the local value
at the receiver point $\boldsymbol{x}$ multiplied by a suitable factor,
which corresponds to the integral of the weight function over the missing
part. Mathematically, this can be described by a weight distribution (generalised function) defined as
\begin{equation}\label{eq:bor1}
\alpha\left(\boldsymbol{x},\boldsymbol{\xi} \right) =
\alpha_\infty(\Vert\boldsymbol{x}-\boldsymbol{\xi} \Vert) +
v(\boldsymbol{x})\delta(\boldsymbol{x}-\boldsymbol{\xi})
\end{equation} 
where $\alpha_\infty$ is the normalised weight function corresponding to
an infinite medium (with no influence of boundaries),
\begin{equation}
v(\boldsymbol{x}) = 1-\int_V \alpha_{\infty}(\Vert\boldsymbol{x}-\boldsymbol{\xi}\Vert){\rm d}\boldsymbol{\xi}
\end{equation} 
is the relative weight of the missing volume, and $\delta$ is the Dirac distribution. Function $\alpha_\infty$ depends only on the distance between points
$\boldsymbol{x}$ and $\boldsymbol{\xi}$ and is normalised such that
the integral of  $\alpha_\infty$ over the whole space (e.g.\ over ${\cal R}^2$ 
in a two-dimensional setting) is equal to unity. Consequently,
in an infinite medium, $v(\boldsymbol{x})=0$.

The physical meaning of formula (\ref{eq:bor1}) is that the nonlocal
variable, e.g.\ nonlocal equivalent strain, is computed as the sum
of the weighted average evaluated with a ``fixed'' weight function
$\alpha_\infty$ (unaffected by the boundary) and an additional term
that contains the local value:
\begin{equation}
\bar{\varepsilon}_{\rm eq}\left(\boldsymbol{x}\right) = \int_{V} \alpha_\infty \left(\Vert\boldsymbol{x}-\boldsymbol{\xi}\Vert\right) \varepsilon_{\rm eq}(\boldsymbol{\xi}) {\rm d}\boldsymbol{\xi} + v(\boldsymbol{x}) \varepsilon_{\rm eq}(\boldsymbol{x})
\end{equation} 
The added term can be called the ``local complement'', hence the name of this approach.

\section{Meso-scale model}\label{sec:mesomodel}
In this work, a meso-scale description of the fracture process in three-point bending tests has been used to create reference results for the evaluation of nonlocal models presented in Section~\ref{sec:macroscopicmodel}.
In this meso-scale approach, aggregates, interfacial transition zones (ITZ) and mortar are modelled as separate phases with different material properties. 
For the mortar and ITZ, a random field of tensile strength and fracture energy is applied.
This meso-scale description has been performed by a lattice approach in combination with a damage mechanics model to describe the mechanical response of the three phases \citep{BolSai98}. 
Since it had been previously used for the determination of fracture process zones of concrete subjected to direct tension \citep{GraJir10}, it is here only briefly recalled.

\begin{figure}
\centering
\begin{tabular}{cc}
\includegraphics[width=4.cm]{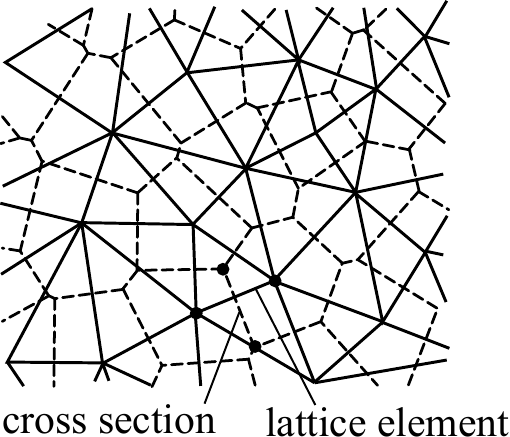} & \includegraphics[width=3.5cm]{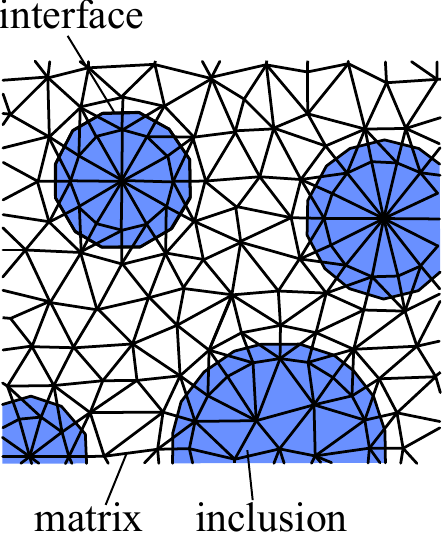}\\
(a) & (b)
\end{tabular}
\caption{(a) Set of lattice elements (solid lines) with middle cross-sections (dashed lines) obtained
from the Voronoi tessellation of the domain; (b) arrangement of lattice elements around inclusions.}
\label{fig:mesoconcept}
\end{figure}

The nodes of the lattice are randomly located in the domain, subject to the constraint of a minimum distance, which is independent of the heterogeneity of the material. 
The lattice elements are obtained from the edges of the triangles of the Delaunay triangulation of the domain (solid lines in Fig.~\ref{fig:mesoconcept}a), whereby the middle cross-sections of the lattice elements are the edges of the polygons of the dual Voronoi tessellation (dashed lines in Fig.~\ref{fig:mesoconcept}a). 

Each lattice node possesses three degrees of freedom, namely two translations and one rotation. 
The degrees of freedom of the lattice nodes are linked to two displacement discontinuities at the centre of the middle cross-section of the element.
The displacement discontinuities are transformed into strains by smearing them out over the distance between the two lattice nodes. 
The strains are related to the stresses by an isotropic damage model describing the constitutive response of ITZ and mortar. 

The spatially varying material properties that originate from the heterogeneity of the material are reflected at two levels. 
Large aggregates are modelled directly by placing lattice nodes at special locations, such that the middle cross-sections of the lattice elements form the boundaries between aggregates and mortar (Fig.~\ref{fig:mesoconcept}b). 
The heterogeneity represented by finer particles is described by autocorrelated random fields of tensile strength and fracture energy, which are assumed to be fully correlated.
The random fields are characterised by an autocorrelation length that is independent of the spacing of lattice nodes. 
Discretely modelled aggregates are assumed to be linear elastic.
This mixed approach, in the form of a discrete representation of the meso-structure and random field, is a compromise between model detail and computational time.
In \cite{GraJir10}, it was shown that the present lattice approach results in crack patterns which are insensitive to the background lattice. 
Furthermore, in the same work it was also shown that the crack openings are independent of the size of the elements.

The heterogeneity introduced in the meso-scale analyses leads to a scatter for the results.
Therefore, all meso-scale results are shown as an average of 100 random analyses. In \cite{GraJir10}, it was shown that this number of analyses is sufficient to give a reliable estimate of the mean.
It should be noted that the computational time required to perform 100 meso-scale analyses is much greater than the one required to perform a single nonlocal analysis. This is one of the main reasons why it is desired to develop a nonlocal approach, which can represent the response obtained by more detailed meso-scale analyses.

\section{Analyses}

\subsection{Calibration based on direct tensile test}

The special scaling approaches (distance-based and stress-based) were implemented into OOFEM \citep{PatBit01,Pat12}, an object-oriented finite element code which provides extensive support for nonlocal simulations.  
Before actually running simulations of beams subjected to bending, the parameters of the nonlocal models were calibrated, so that the model results for direct tensile loading were in agreement with meso-scale results reported in \cite{GraJir10}. 

The idea was to determine the basic parameters in a simple test in which the influence of boundaries is eliminated.
In a direct tensile test, the stress is uniform and the strain remains uniform until the onset of localisation.
The position of the localised process zone depends on random imperfections and can be selected 
(e.g.\ by placing a slightly weakened element in the middle of the specimen)
such that the process zone is 
unaffected by the boundaries, of course provided that the specimen is sufficiently long. A similar approach had already been used
 in \cite{GraJir10} to show that an excellent agreement in terms of the load-displacement curve can be obtained with 
different sets of parameters, involving different values of the internal length $R$. This ambiguity was removed by comparing not only the load-displacement curves but also the profiles of dissipated energy density.

The nonlocal simulations in \cite{GraJir10} were done in the one-dimensional setting, which was perfectly justified
in the context of that paper. However, the objective of the present calibration procedure is to determine parameters
that can be used in two-dimensional simulations of bending failure. Therefore, the nonlocal averaging at points unaffected
by the boundaries must be done using the same averaging scheme in the calibration on the direct tensile test
as in the subsequent application to bending beams. To achieve that, one should simulate the tensile specimen using 
a two-dimensional finite element mesh with periodicity conditions imposed on the boundaries parallel to the direction
of loading. In fact, the periodicity conditions can be replaced by constraints enforcing the transversal displacements
(perpendicular to the direction of loading) to vanish on one of the boundary lines  parallel to the direction of loading and to have equal values on the other boundary line parallel to the direction of loading. 
This is perfectly consistent with the conditions imposed in the mesoscale simulations in \cite{GraJir10}. In the nonlocal simulation, the transversal strain is uniform in the entire ``periodic cell'' and the resulting transversal stress becomes nonzero after the onset of localisation.
Of course, it vanishes on the average, but its local value varies as a function of the axial coordinate and typically is positive in the localised process zone and negative in the elastically unloading zone. 
Therefore, the test cannot be interpreted as a typical tensile test on a narrow bar (in which necking in the process zone is possible) but rather as a tensile test of a plate which has an infinite dimension in the transversal direction. Let us emphasise again that this setup perfectly corresponds to the conditions imposed in the mesoscale analysis.

To reduce the computational effort, the simulation can be done on a single layer of two-dimensional elements with one integration point per element and with a modified nonlocal weight function. Indeed, if the two-dimensional averaging with weight function $\alpha_\infty$ is applied on a local field that depends on one spatial coordinate only, say $x_1$, and is constant in the direction of the other coordinate axis, say $x_2$, one gets
\begin{eqnarray}\label{eq:nonlocalDamageXY}
\bar{\varepsilon}_{\rm eq}\left(x_1,x_2\right) = \int_{-\infty}^{\infty}\int_{-\infty}^{\infty} \alpha_\infty \left(\sqrt{(x_1-\xi_1)^2+(x_2-\xi_2)^2}\right) \varepsilon_{\rm eq}(\xi_1) \rm{d}\xi_1\rm{d}\xi_2 = 
\int_{-\infty}^{\infty} \alpha_\infty^* \left(x_1-\xi_1\right) \varepsilon_{\rm eq}(\xi_1) \rm{d}\xi_1
\end{eqnarray} 
where
\begin{equation}\label{eq:nonlocalDamageX}
\alpha_\infty^*(x) = \int_{-\infty}^{\infty}\alpha_\infty \left(\sqrt{x^2+s^2}\right)\rm{d}s
\end{equation}
is a modified weight function (reduced from 2D to 1D). Interestingly, if the original weight function $\alpha_\infty$ is Gaussian,
it is not affected by the reduction of spatial dimensions and $\alpha_\infty^*$ remains Gaussian. 
In the present work we use a non-Gaussian two-dimensional
weight function obtained by normalising the function defined in (\ref{eq:basicAlpha0}). 
The corresponding one-dimensional weight function $\alpha_\infty^*$ cannot be expressed in closed form,
but it can be evaluated numerically from (\ref{eq:nonlocalDamageX}). After this adjustment, the analysis can be run efficiently
with the same number of degrees of freedom as in the one-dimensional case (except for one additional degree of freedom
for the lateral displacement) and with results fully equivalent to what would be obtained by the two-dimensional averaging scheme.

Nonlocal simulations were run on a specimen of length $L=100$ mm and the resulting load-displacement diagrams were converted
into stress-strain diagrams with the average strain (defined as the change of specimen length divided by the initial length)
 plotted on the horizontal axis. The parameters were modified until the best possible agreement with the corresponding
diagram obtained for the mesoscale model was achieved. The characteristic length $R$ was adjusted to get a reasonable
agreement with the profile of dissipated energy density along the specimen.  
The calibration resulted in elastic parameters of $E = 29.6$~GPa and $\nu=0.2$, which are the same for the damage model and the damage-plasticity model. 
The other parameters of the damage model are $f_{\rm t} = 2.86$~MPa, $\varepsilon_{\rm{max}} = 0.000198$, $\varepsilon_{\rm{1}} = 0.00024$, $\varepsilon_{\rm{2}} = 0.00052$, $n=0.85$ and $R = 4$~mm.
For the damage-plasticity model, two values of the parameter $m$ were used, namely $m=1$ and $m=2$. The value of $m=1$ corresponds to the standard nonlocal averaging, whereas $m =2$ gives an over-nonlocal model.
The parameters of the plastic part of the damage-plasticity model control the pre-peak part of the load-displacement curve and their optimised values are $f_{\rm t} = 2.86$~MPa, $\varepsilon_{\rm {p,max}} = 0.0001234$, $E_0 = 1480$~GPa.
Optimised parameters of the damage part of the damage-plasticity model, which control softening, were found to be $d_1 = 0.08 $, $d_2 = 1.3$, $d_3 = 0.04$, $d_4 = 1$ and $d_5= 0.65$ and $R = 5$~mm for  $m = 1$,  and $d_1 = 0.08$, $d_2 = 1.3$, $d_3 = 0.08$, $d_4 = 0.9$ and $d_5 = 0.6$ and $R = 2.4$~mm for $m = 2$.

\begin{figure}
\begin{center}
\begin{tabular}{cc}
\includegraphics[width=9cm]{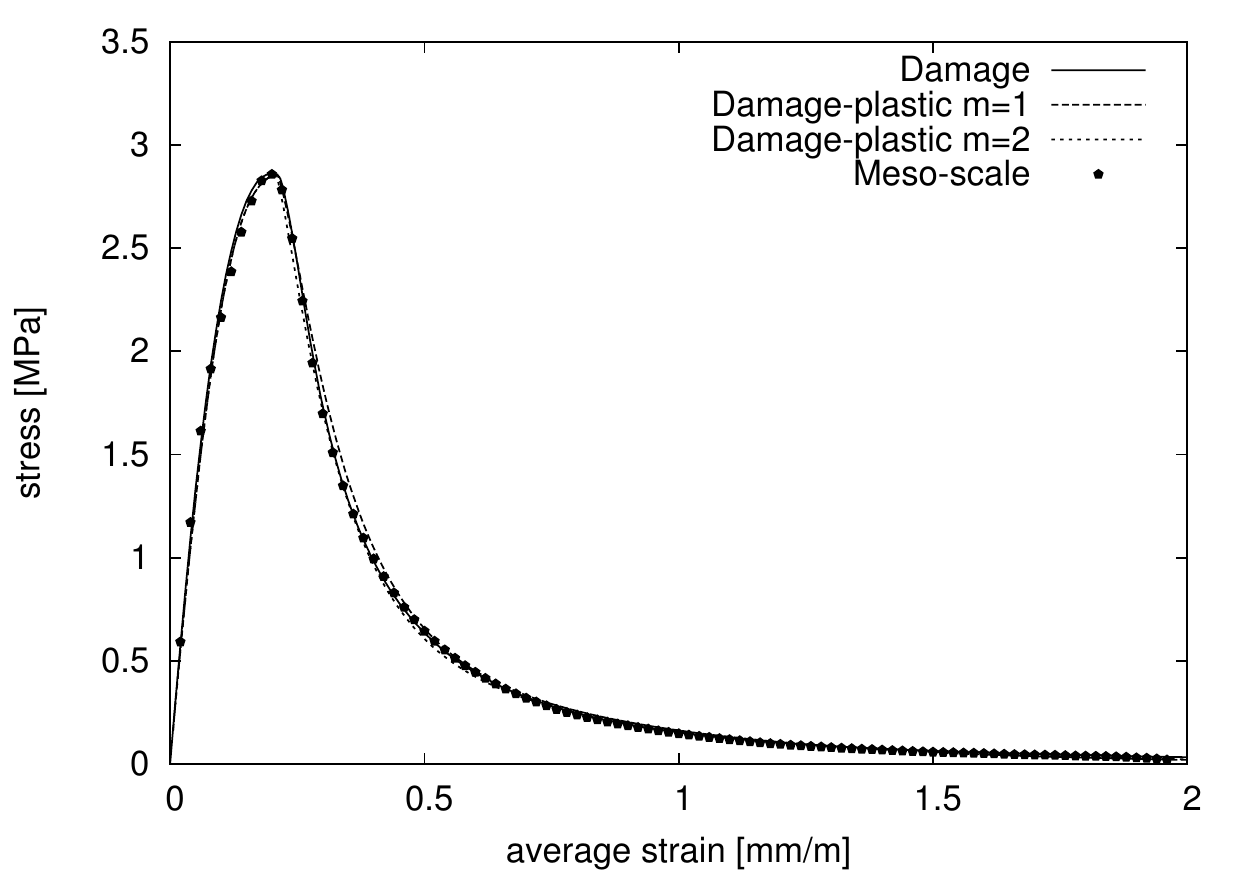} & \includegraphics[width=9cm]{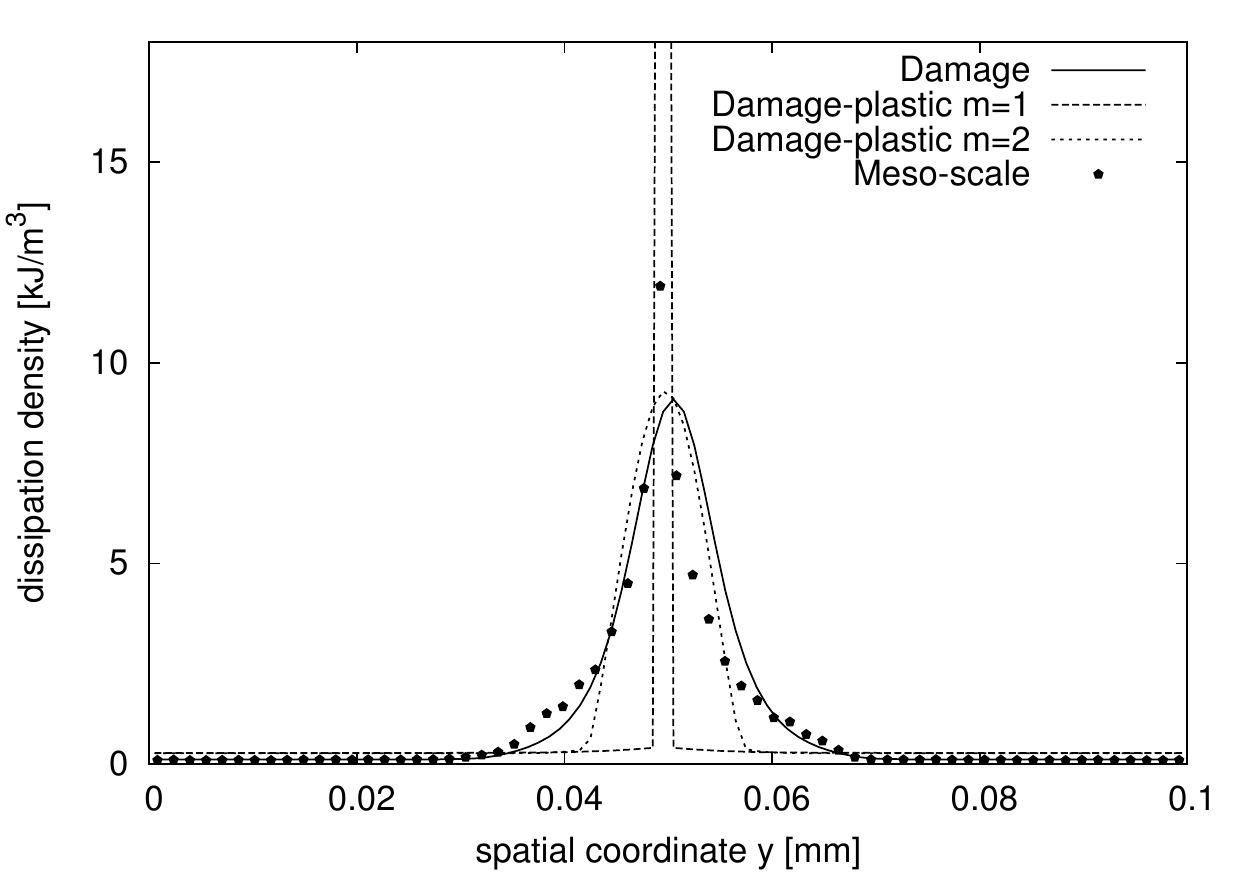}\\
(a) & (b)
\end{tabular}
\end{center}
\caption{1D calibration: (a) stress-strain curve, (b) dissipated energy density across the fracture process zone in the last loading step.}
\label{fig:1danalysis}
\end{figure}

The results of the calibration are shown in Fig.~\ref{fig:1danalysis} in the form of stress-average strain curves and dissipated energy density across the fracture process zone. 
The stress-strain curves of the nonlocal models agree very well with the meso-scale results.
The dissipation profile of the nonlocal damage model fits the curve obtained from meso-scale analyses quite well.
For the over-nonlocal damage-plasticity model ($m=2$), the shape of the profile is somewhat different but the width of the process zone can be adjusted to get the best fit in the least-square sense. For the standard nonlocal formulation ($m=1$), plastic strain localises into one element and since the dissipation in the damage zone around the fully localised plastic zone is very small, the shape of the dissipation profile cannot be captured properly. Still, the total dissipation is enforced to be the same as in the meso-scale analysis, and the load-displacement curve does not exhibit pathological sensitivity to the element size.
The meso-scale results in Fig.~\ref{fig:1danalysis} represent an average of 100 analyses.

\subsection{Evaluation of boundary effects based on bending tests}
\begin{figure}
\centering
\includegraphics[width=8cm]{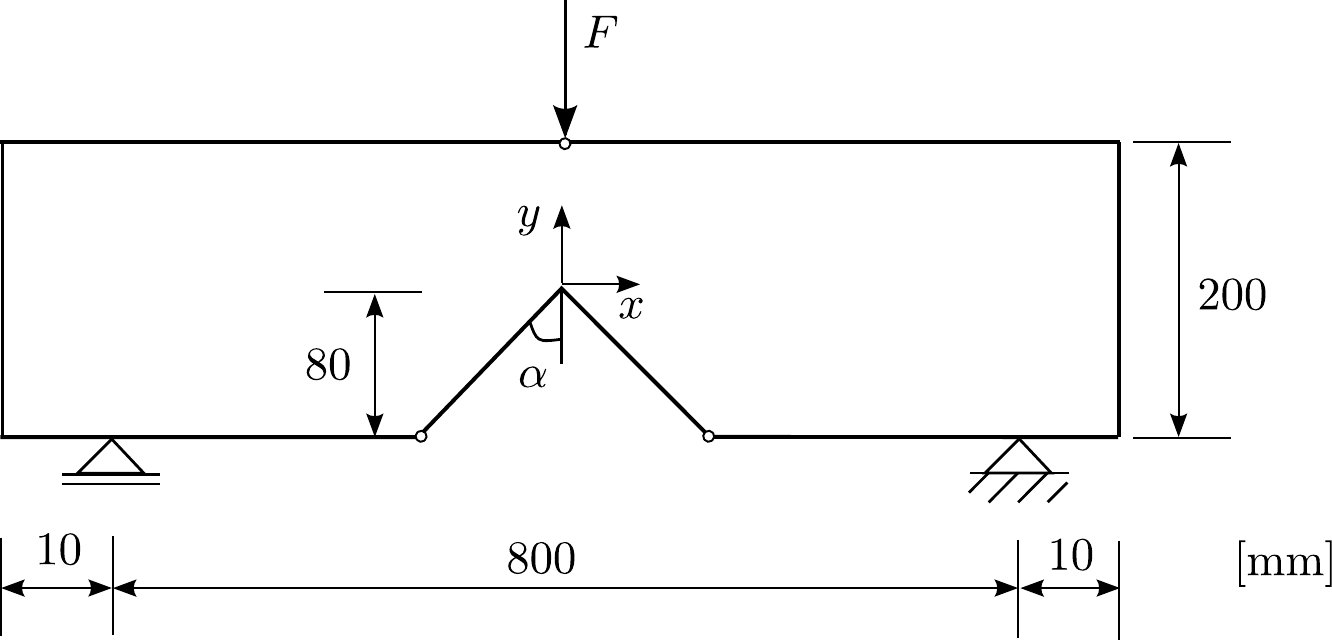}
\caption{Geometry and loading setup of the notched beams subjected to three point bending.}
\label{fig:beamgeo}
\end{figure}

In the second step, the nonlocal models combined with the different boundary approaches were applied to 2D plane-stress analyses of notched three-point bending tests. 
The geometry and the loading setup are shown in Fig.~\ref{fig:beamgeo}. 
Three boundaries in the form of a sharp notch with $\alpha=0^{\circ}$, a V-notch with $\alpha=45^{\circ}$ and an unnotched boundary with $\alpha = 90^{\circ}$ were considered. 
The boundary types were chosen so that the performance of the boundary approaches (Section~\ref{sec:boundaries}) could be compared.

\subsubsection{Nonlocal damage models}

\begin{figure}
\begin{center}
\begin{tabular}{cc}
\includegraphics[width=9cm]{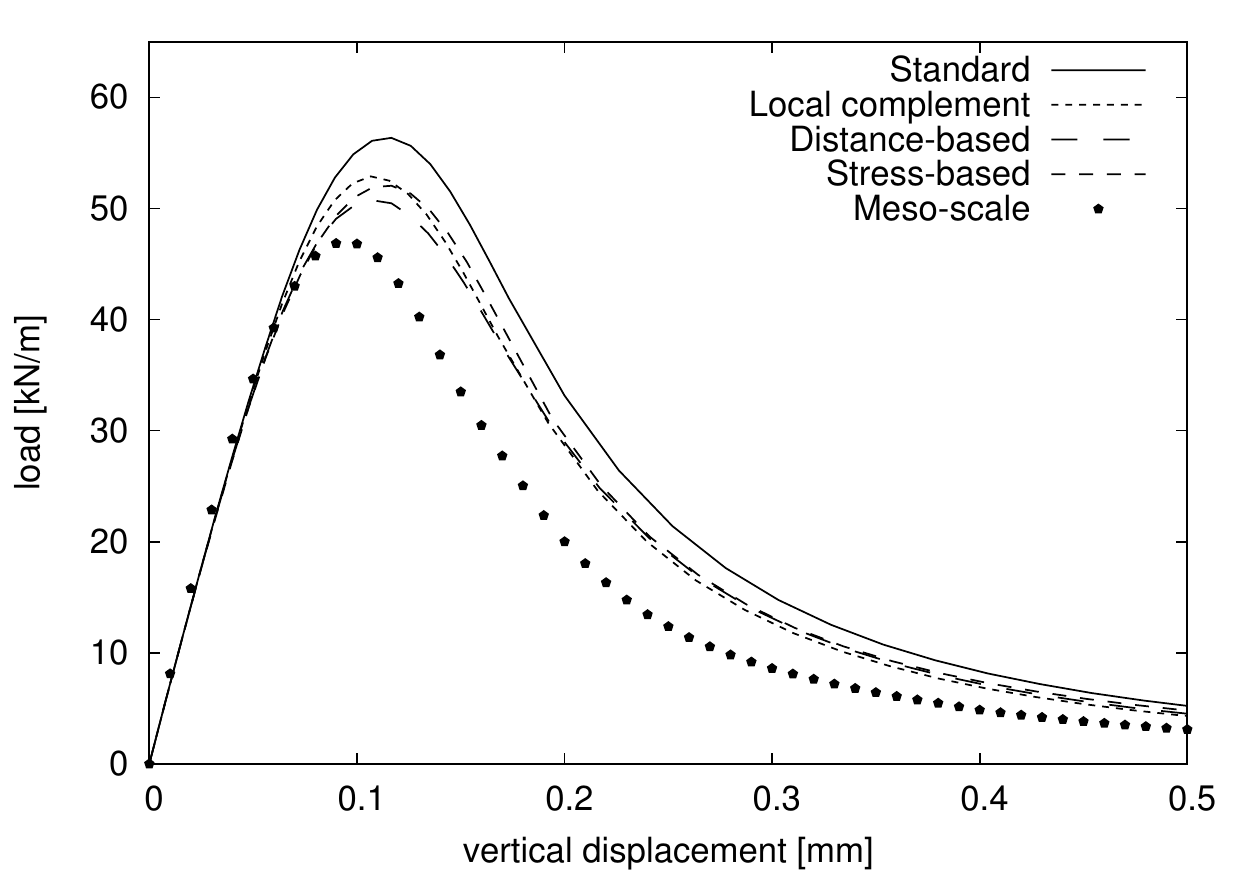}  & \includegraphics[width=9cm]{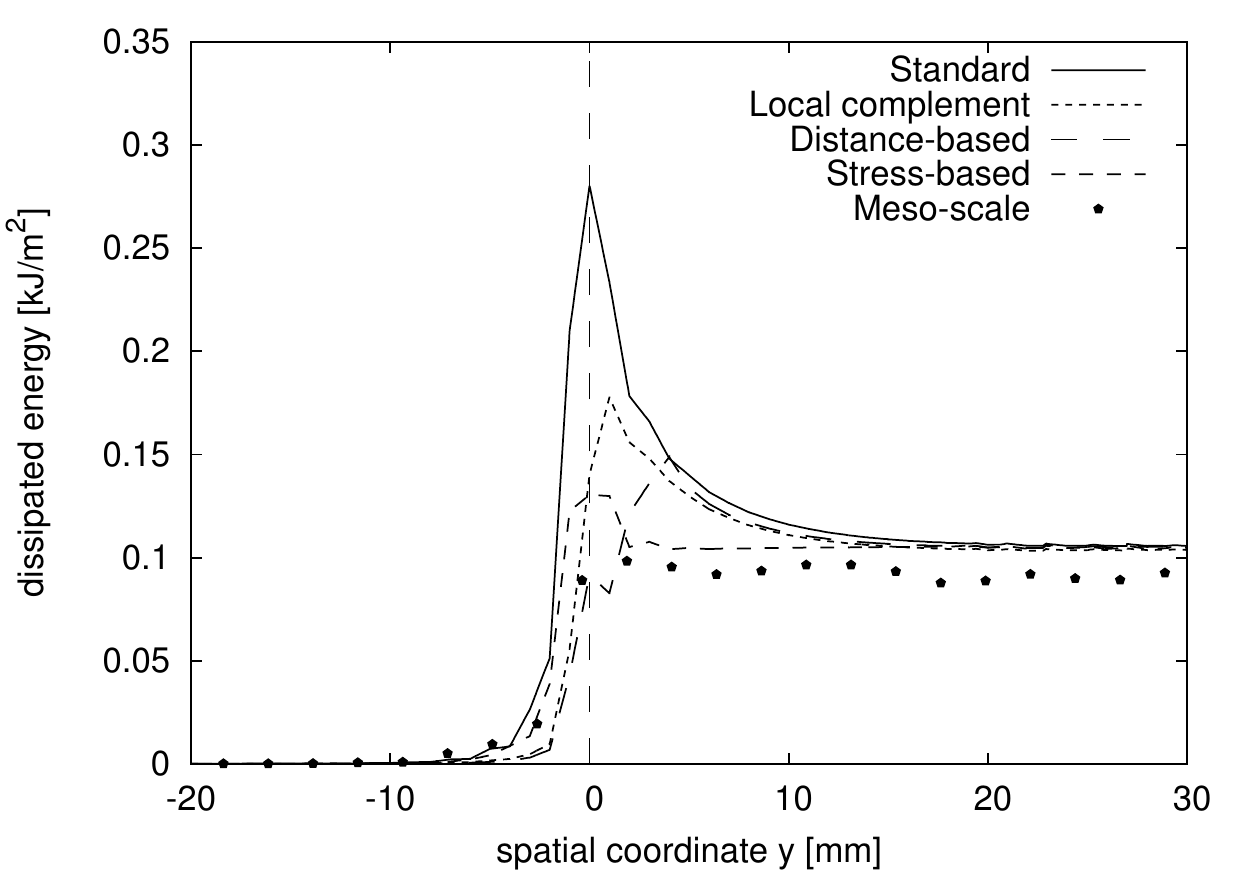}\\
(a) & (b)
\end{tabular}
\end{center}
\caption{Comparison of the results of four {\bf nonlocal damage} approaches and meso-scale analysis for specimen with a {\bf sharp notch} ($\alpha = 0^{\circ}$): (a) load-displacement curves and (b) dissipated energy profiles.}
\label{fig:2d0a}
\end{figure}

\begin{figure}
\begin{center}
\begin{tabular}{cc}
\includegraphics[width=9cm]{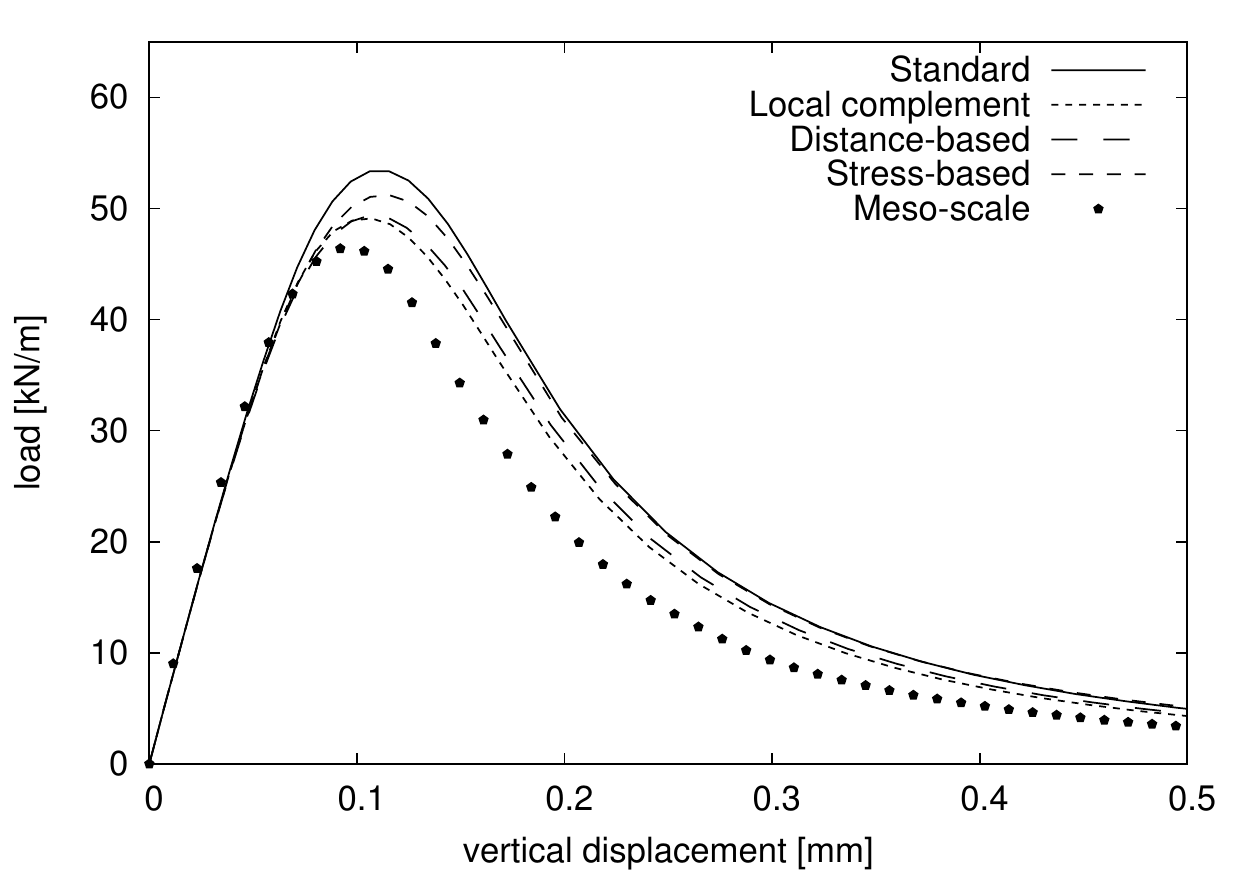}  & \includegraphics[width=9cm]{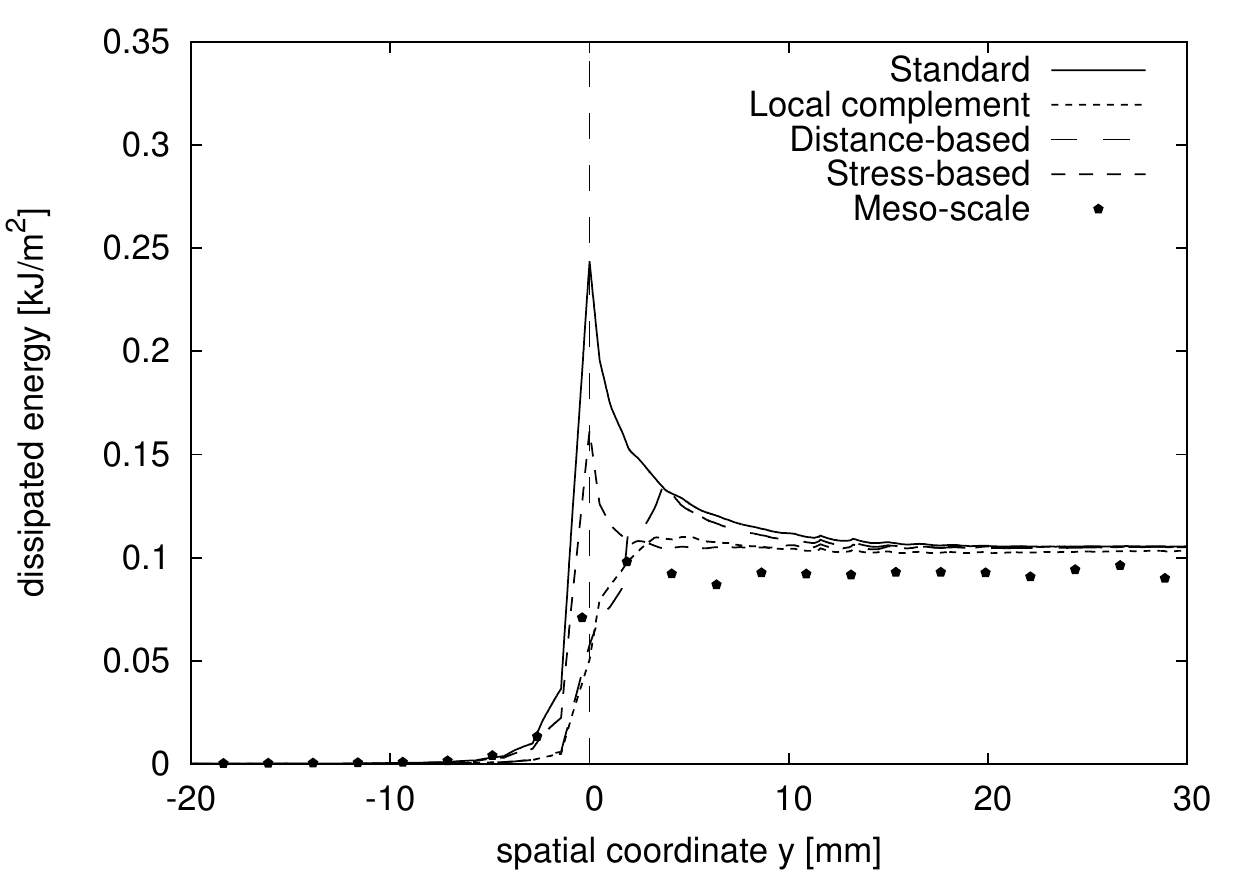}\\
(a) & (b)\\
\end{tabular}
\end{center}
\caption{Comparison of the results of four {\bf nonlocal damage} approaches and meso-scale analysis for specimen with a {\bf V-notch} ($\alpha = 45^{\circ}$): (a) load-displacement curves and (b) dissipated energy profiles.}
\label{fig:2d45a}
\end{figure}

\begin{figure}
\begin{center}
\begin{tabular}{cc}
\includegraphics[width=9cm]{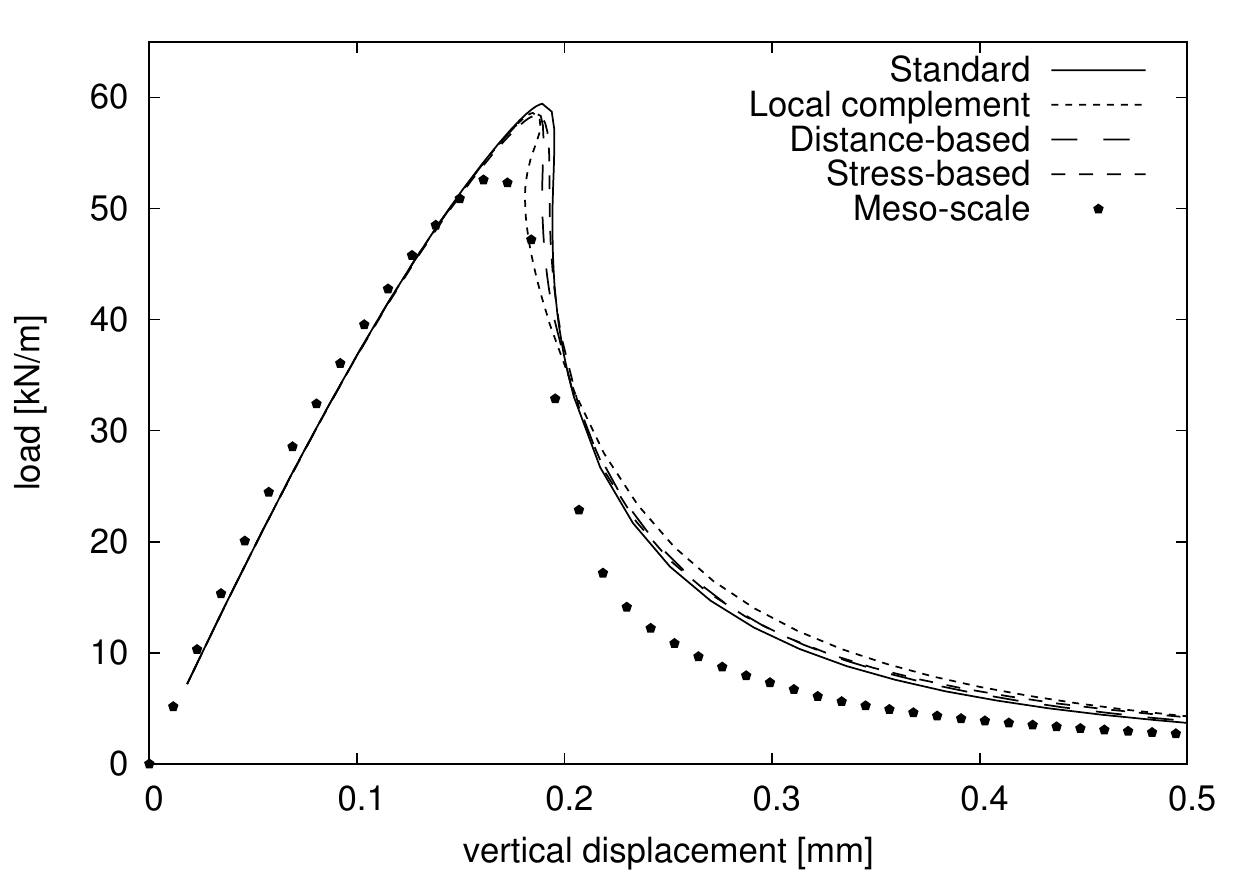} & \includegraphics[width=9cm]{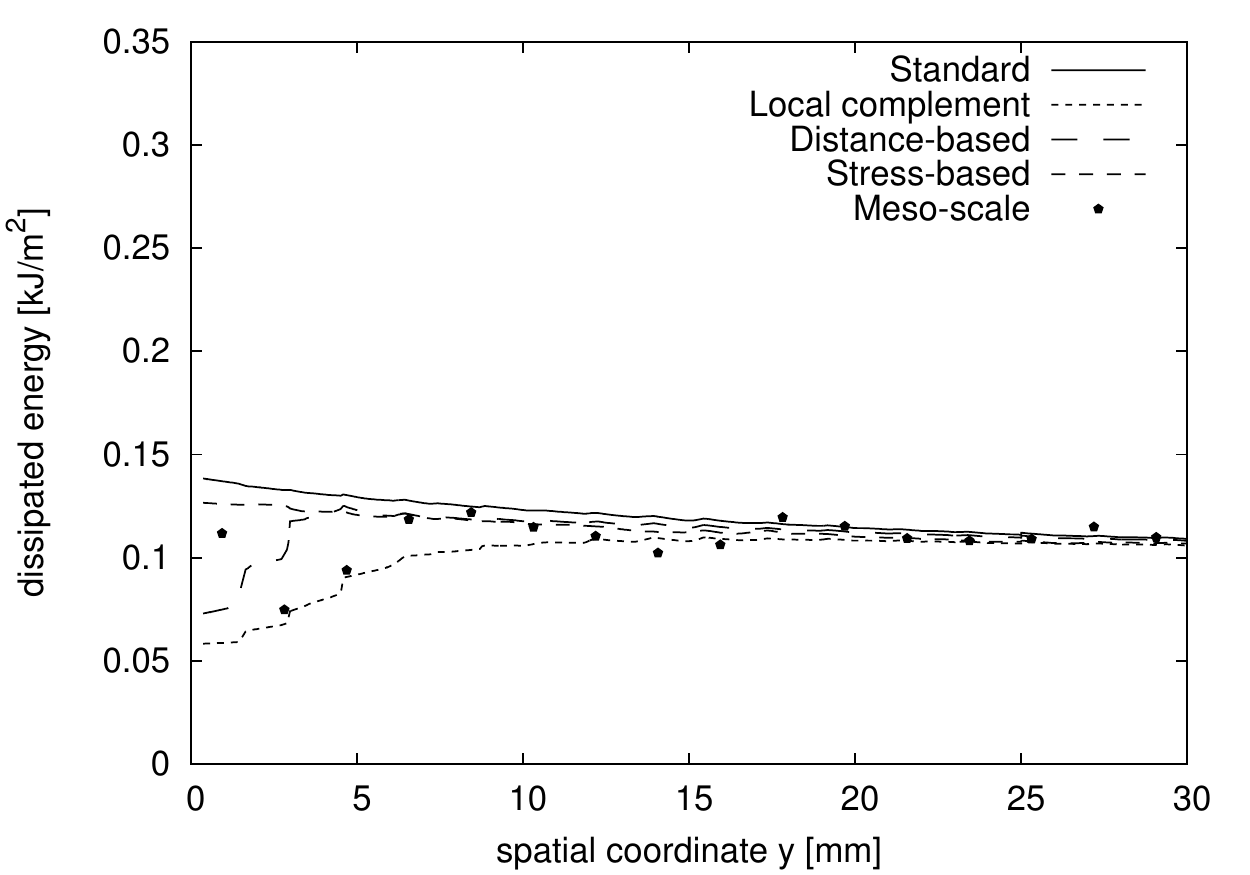}\\
(a) & (b)\\
\end{tabular}
\end{center}
\caption{Comparison of the results of four {\bf nonlocal damage} approaches and meso-scale analysis for the {\bf unnotched} specimen ($\alpha = 90^{\circ}$): (a) load-displacement curves and (b) dissipated energy profiles.}
\label{fig:2d90a}
\end{figure}

The additional parameters needed for the distance-based modification of averaging near the boundary were chosen as $\beta = 0.15$ and $t = 1$, and for the stress-based modification as $\beta=0.15$.
The results obtained with the nonlocal damage approaches for the three beam geometries are compared to meso-scale results in Figs.~\ref{fig:2d0a}--\ref{fig:2d90a} in the form of load-displacement curves and dissipated energy distribution along the ligament of the beam. 
The dissipated energy per unit length squared was obtained by integrating the dissipation density along the width of the process zone.
For improving the clarity of the figures, the dissipated energy distribution is only shown for the first $3$~cm of the ligament length, for which the notch types are expected to have a strong influence. Same as for the 1D calibration, the meso-scale results are an average of 100 analyses. The standard deviations for the peak load in the meso-scale analyses for $\alpha=0$, $45$ and $90$ are $2.6$, $2$ and $1.3$~kN, respectively.

Let us emphasise that our main objective is not to obtain a perfect fit in terms of the load-displacement diagram, but to detect formulations that do not lead to spurious effects near the boundary (or at least reduce such effects).
As seen in parts (a) of the figures, the peak load of the meso-scale analysis is overestimated by the nonlocal damage model with standard scaling for all beam geometries.
For $\alpha=0^{\circ}$ and 45$^{\circ}$ (notched specimens), this overestimation is accompanied by a much higher dissipation near the notch than in the meso-scale analyses; see Figs.~\ref{fig:2d0a}b~and~\ref{fig:2d45a}b.
The dissipated energy profiles for the stress-based and the distance-based nonlocal damage approaches, as well as for the local complement approach, are in much better agreement with the meso-scale results for these beam geometries.
In particular, the energy peaks close to the notch that arise for the standard approach are removed, or at least substantially reduced.

For most approaches, farther from the notch, the dissipated energy is more or less uniformly distributed along the ligament but, for the nonlocal damage approaches, its value is slightly overestimated compared to the meso-scale results. This effect may be related  to the multiaxiality of the stress state in the bending test, as discussed in a different context by \cite{JirBau12}, and it could be reduced by a different choice of the equivalent strain expression. It is sometimes overlooked that bending may lead to nonnegligible stresses in the direction perpendicular to the ``beam fibers'', and thus to biaxial stress states. This effect is especially strong near a notch, which is documented by the fact that the singular part of the mode-I stress field computed according to linear elastic fracture mechanics is highly biaxial. An illustration of this phenomenon can be found in Figs.~12, 13 and 15 of \cite{JirBau12}.

The stress-based approach leads to a somewhat lower dissipation density and is thus closer to the results of meso-scale analyses. Furthermore, this approach requires only one additional parameter.

For $\alpha=90^{\circ}$ (unnotched specimen) all models overestimate the peak load, but they do not lead to dramatic energy peaks near the boundary; see Fig.~\ref{fig:2d90a}b.
However, the standard and stress-based damage model still overestimate the meso-scale results near the boundary. On the other hand, the local complement damage model underestimates the dissipation obtained with the meso-scale model.

Concerning the performance of the local complement model for the three notch types, comparing the energy dissipation profiles, one can observe that the bigger the area outside the beam the smaller is the dissipated energy. Thus, for the local complement model, the performance is strongly influenced by the geometry of the notch. For a sharp notch, some excessive dissipation is still observed, albeit lower than for the standard averaging. For a V-notch, the results agree with the meso-scale analysis very well, while for a beam without a notch the
dissipation density in the boundary layer is somewhat too low.  

\begin{figure}
\begin{center}
\begin{tabular}{cc}
\includegraphics[width=9cm]{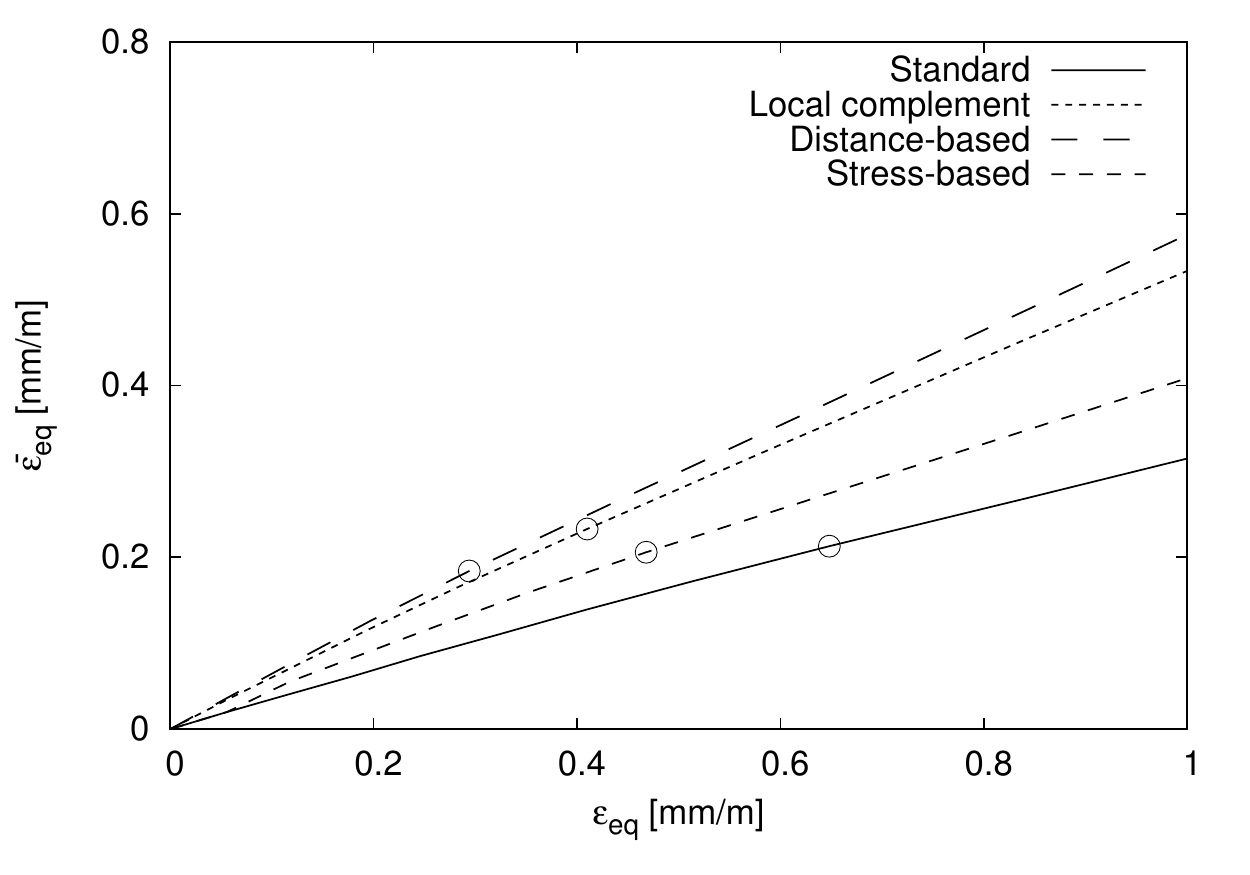}  & \includegraphics[width=9cm]{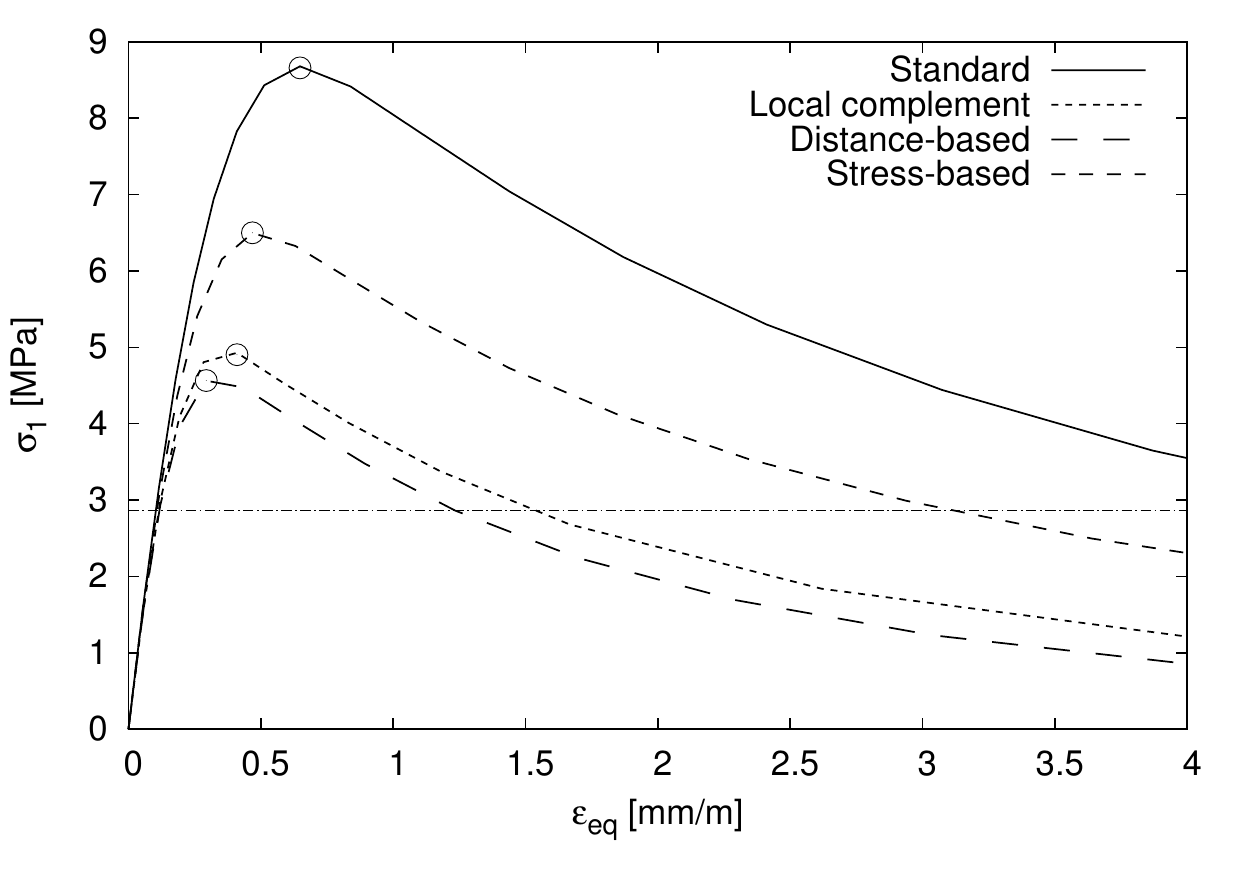}\\
(a) & (b)
\end{tabular}
\end{center}
\caption{Response of an integration point directly above the V-shaped notch for the four nonlocal damage models: (a) nonlocal versus local equivalent strain, (b) maximum principal stress versus local equivalent strain. The circles in (a) and (b) indicate the state for which the contour plot of local equivalent strain in Fig.~\ref{fig:contour} is shown.}
\label{fig:2dratio45}
\end{figure}

The differences in the responses of the four damage models are further illustrated by studying the internal variables at a selected integration point located directly above the notch for the V-notched specimen, which represents an intermediate case between the sharply notched and the unnotched beams. 
In Fig.~\ref{fig:2dratio45}a, the nonlocal equivalent strain versus the local equivalent strain is shown.
The nonlocal equivalent strain is used to calculate the damage variable in (\ref{eq:damageDamage}) and is itself determined by a weighted average, which is performed over contributing regions of different shapes and sizes depending on the approach used; see Section~\ref{sec:macroscopicmodel}.
In the limit of a contributing region of zero radius, the nonlocal equivalent strain determined by the weighted average would be equal to the local equivalent strain. 
The actual averaging uses a nonzero interaction radius, and the nonlocal strain at points near the centre of the process zone is reduced, due to the proximity of regions with lower local strains. 
At a point directly above the notch tip, this is the case, since this point is highly strained whereas material in its vicinity is subjected to smaller strains. 
The reduction of the nonlocal equivalent strain leads to a decrease of the damage and therewith to an increase of the maximum stress reached at this point, which explains the excessive energy dissipation for some of the nonlocal models.

\begin{figure}
\begin{center}
\begin{tabular}{ccccc}
\includegraphics[width=3cm]{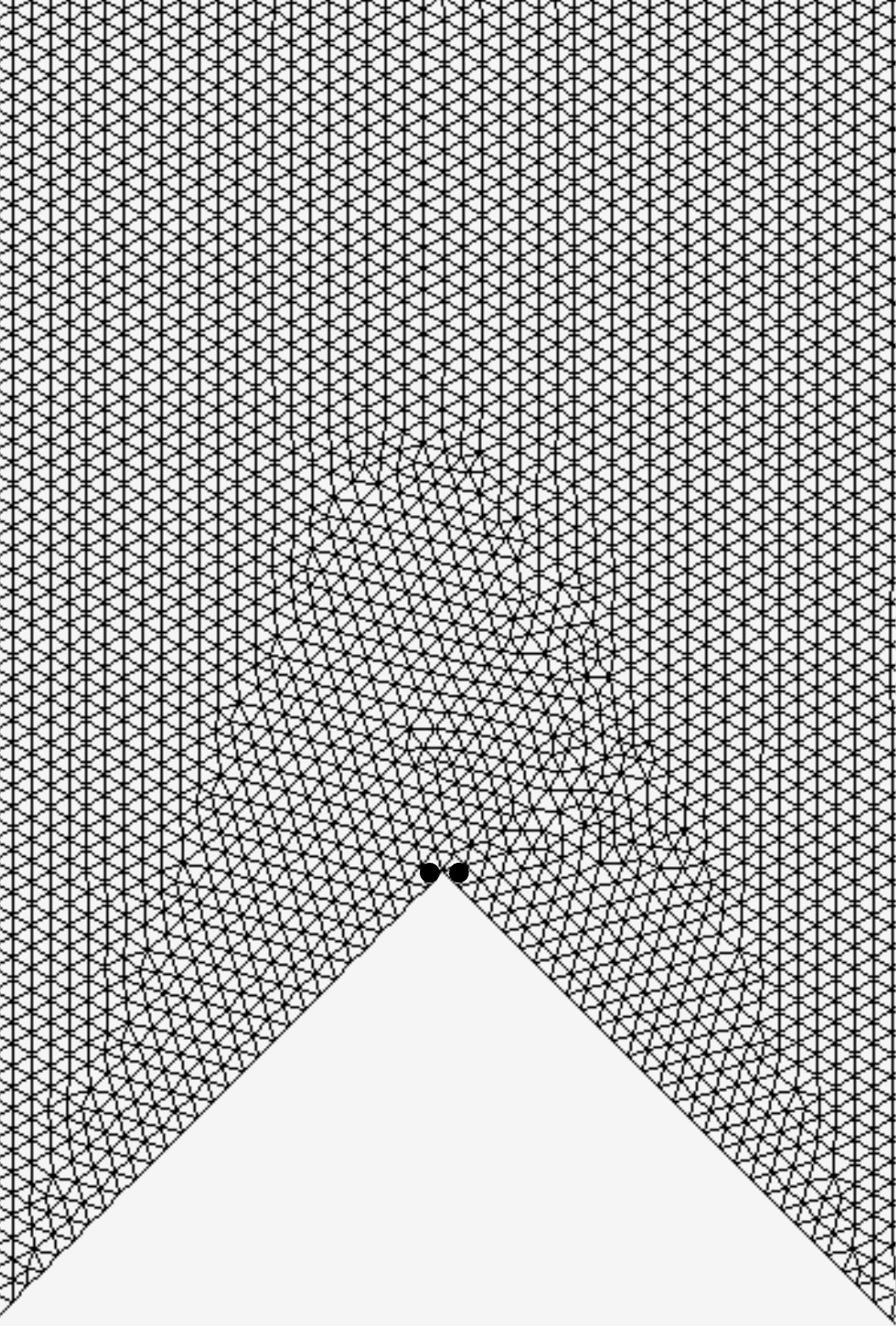} & \includegraphics[width=3cm]{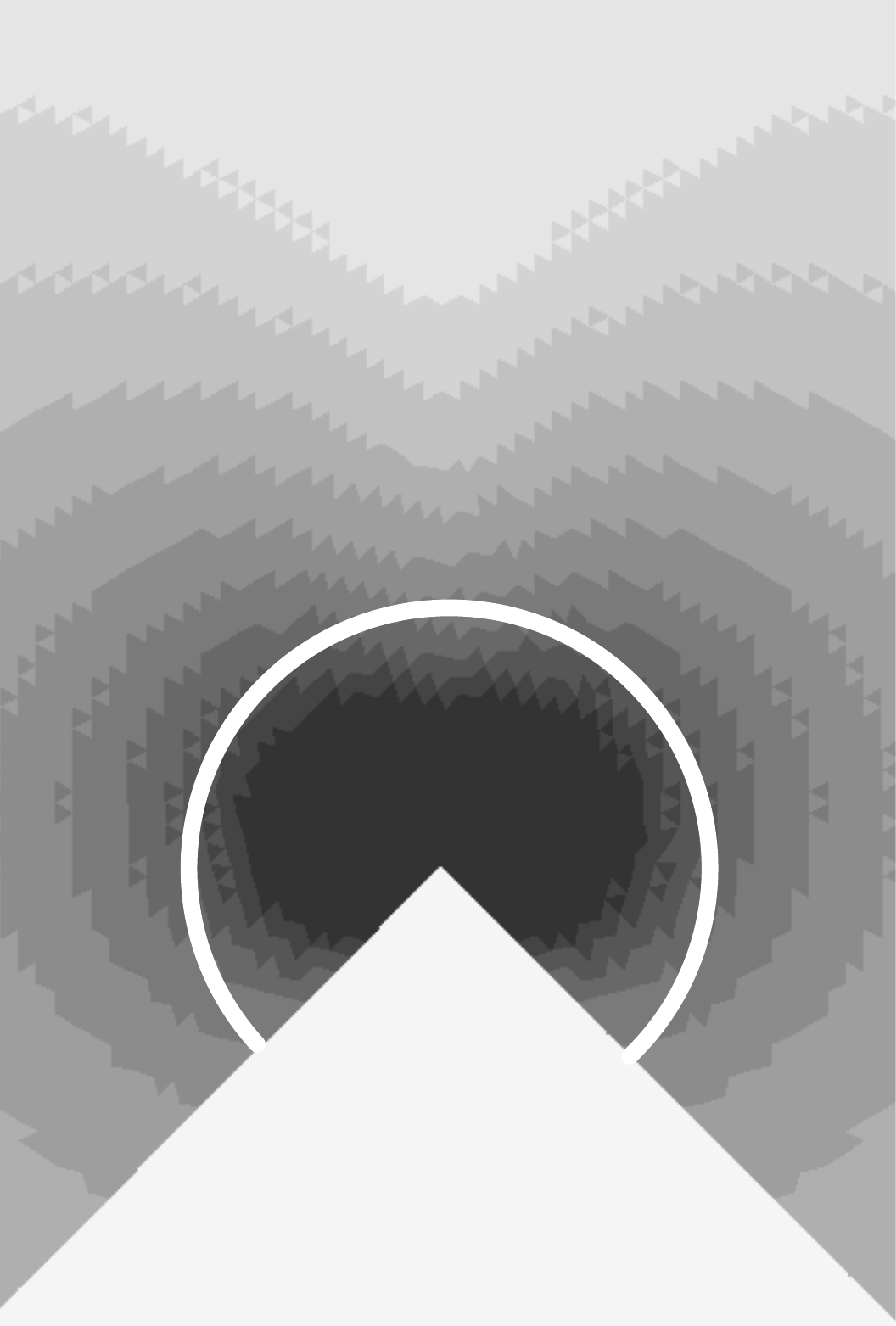} & \includegraphics[width=3cm]{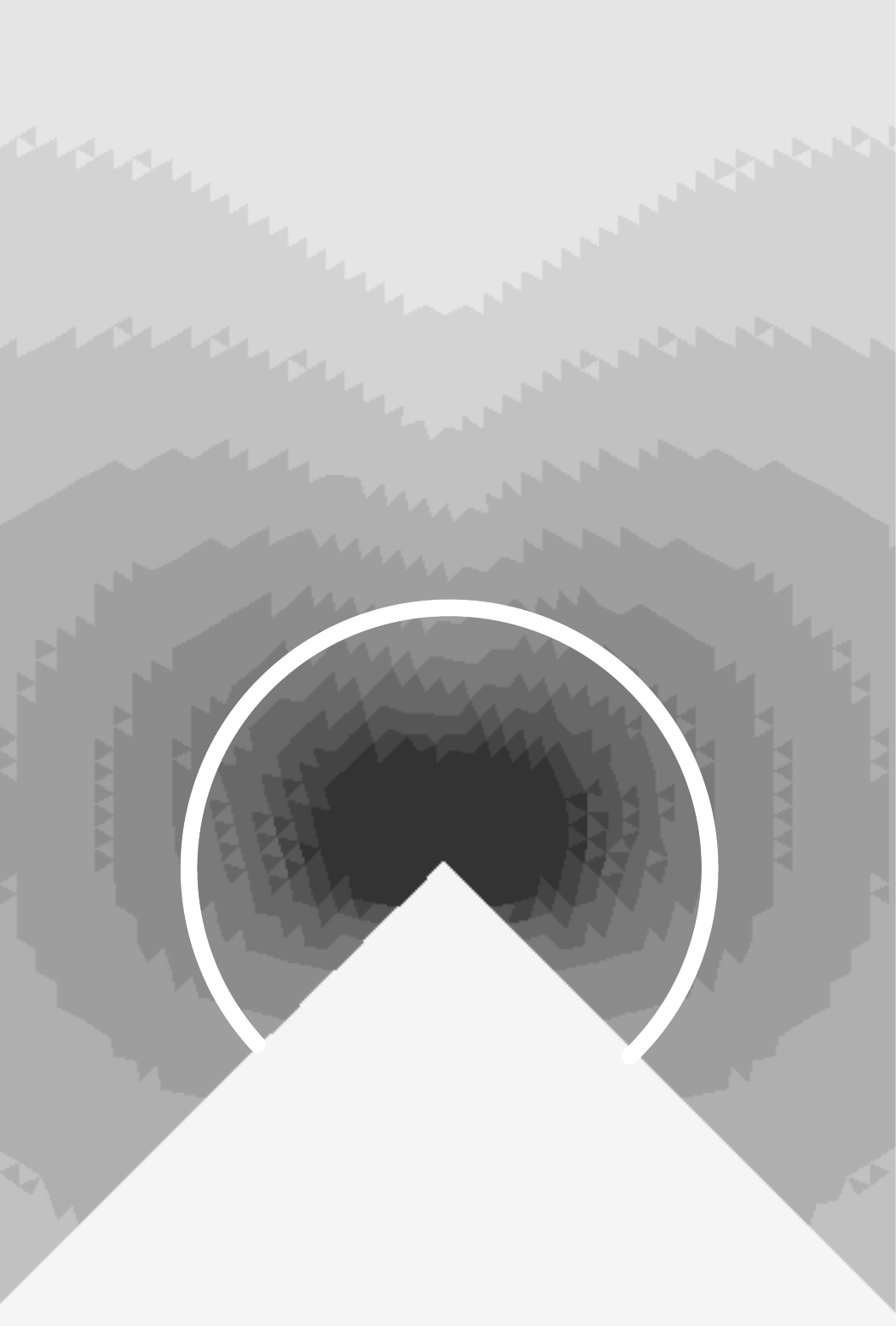} & \includegraphics[width=3cm]{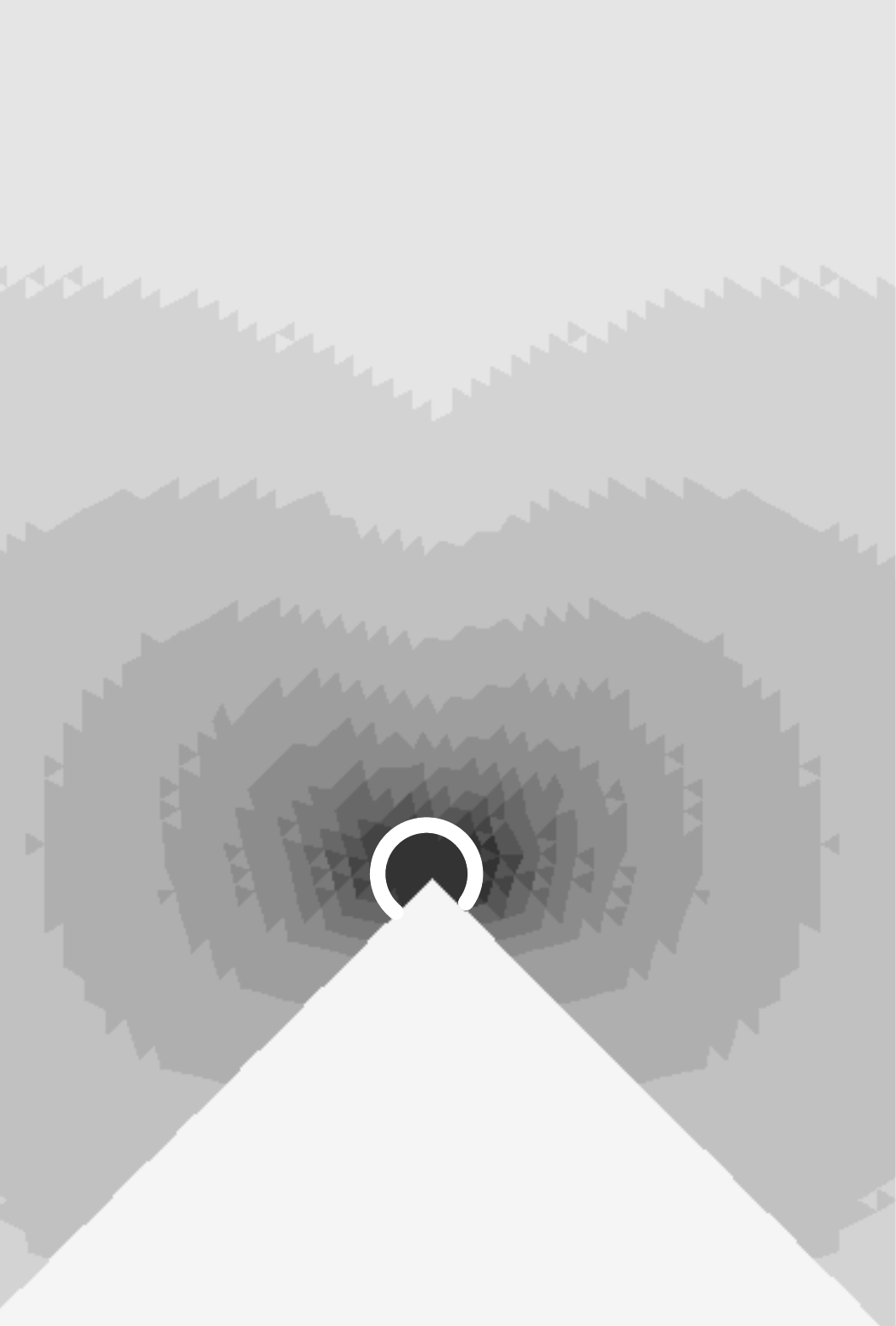} & \includegraphics[width=3cm]{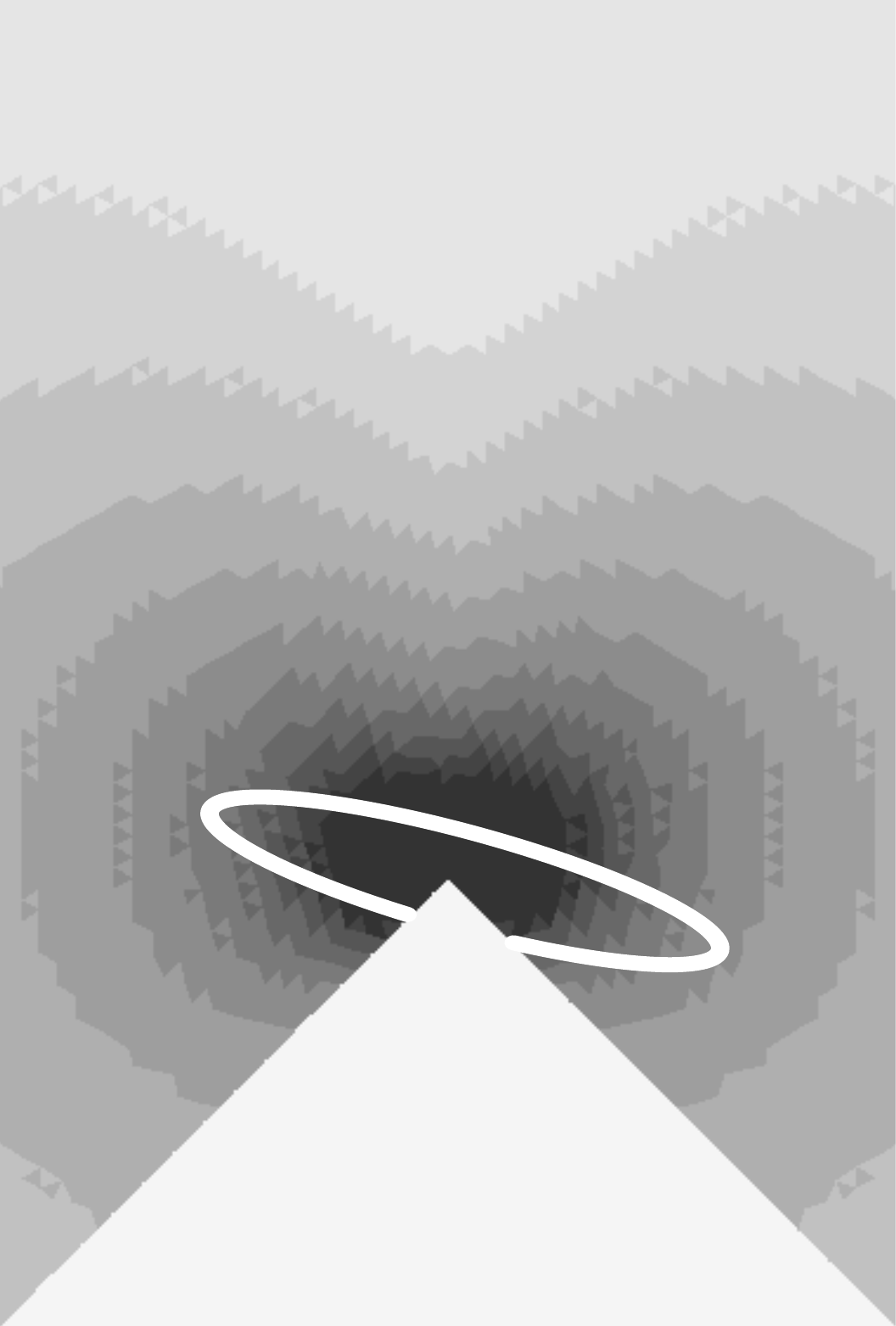}\\
(a) & (b) & (c) & (d) & (e)\\
\end{tabular}
\end{center}
\caption{(a) Finite element mesh near the V-notch, (b)--(e) contour plots of the local equivalent strain for the four damage models:  (b) standard averaging, (c) local complement, (d) distance-based approach and (e) stress-based approach, plotted for the states marked in Fig.~\ref{fig:2dratio45}. The white shapes indicate the boundaries of the region that contributes to the weighted average of equivalent strain at the integration point near the notch considered in Fig.~\ref{fig:2dratio45}. The black colour indicates equivalent strain levels of 0.0001 and greater. Only a part of the beam depth is shown.}
\label{fig:contour}
\end{figure}

As seen in Fig.~\ref{fig:2dratio45}a, for all four damage models the nonlocal strain is smaller than for the local one, with the distance, local complement and stress-based models exhibiting smaller reductions than the standard model.
For the standard, local complement and stress-based damage models, an element slightly to the right of the notch tip was selected, since for these three approaches the strains were concentrated in this element (Fig.~\ref{fig:contour}a). However, for the distance-based damage model, the strains were concentrated in an element slightly on the left of the notch tip. Therefore, this element was selected for the distance-based damage model.
In Figs.~\ref{fig:contour}b-e, the contour plots of the local equivalent strain for the four damage models show that the distance-based and local complement models give smaller zones of high values of local equivalent strain above the notch. 
Note that the maximum principal stress can attain values that substantially exceed the ``tensile strength'' $f_{\rm t}=2.86$ MPa. The peak stress would be equal to the tensile strength if the damage and effective stress were both computed from the local strain. However, nonlocal damage models of the kind considered here evaluate the effective stress from the local strain but the damage from the nonlocal equivalent strain. Consequently, the damage near the centre of the process zone just above the notch ``lags behind'' the effective stress, and the nominal stress rises to very high levels. This effect is responsible for the excessive dissipation near the notch and is especially strong for the standard nonlocal averaging; see Fig.~\ref{fig:2dratio45}b.

Of course, the nonlocal equivalent strain at the ``most strained point'' always exceeds the local equivalent strain, even in the simple tensile test on which the basic parameters of the nonlocal model were calibrated. One may wonder why the effect of artificial strength increase was not felt in that case.
There are two closely related reasons for that: (i) In the tensile test, the strain is uniform up to the onset of localisation, which occurs at peak stress. Under uniform strain in the pre-peak range, the nonlocal strain remains equal to the local one, and so the slower growth of nonlocal strain (as compared to the local one) applies to increments from the peak state only, not to the total values. (ii) In the tensile test, the largest values of local strain are found on the entire line perpendicular to the direction of loading, so only the variation of local strain in the direction of loading contributes to the reduction of the nonlocal strain.
In contrast to that, near a crack tip the local strain decreases in all directions from the tip, and the strain near the crack faces below the notch is very small, so the resulting contrast between local and nonlocal strain is much more pronounced.

It is interesting to note that the standard type of nonlocal damage model with a continuous weight function does not actually remove the stress singularity at the tip of a notch. Before the onset of damage, the solution corresponds to linear elastic fracture mechanics. The nonlocal strain is bounded and before its maximum value attains the damage threshold, no damage is induced. The stress is thus initially computed from the local strain using an elastic law and is unbounded. Similar arguments apply to the V-notch, even though the type of singularity is different but the stresses are still unbounded.
The value of the peak stress seen in Fig.~\ref{fig:2dratio45}b depends on the distance of the Gauss point from the notch and could be arbitrarily large if the examined point approaches the notch tip.

\begin{figure}
\begin{center}
\includegraphics[width=9cm]{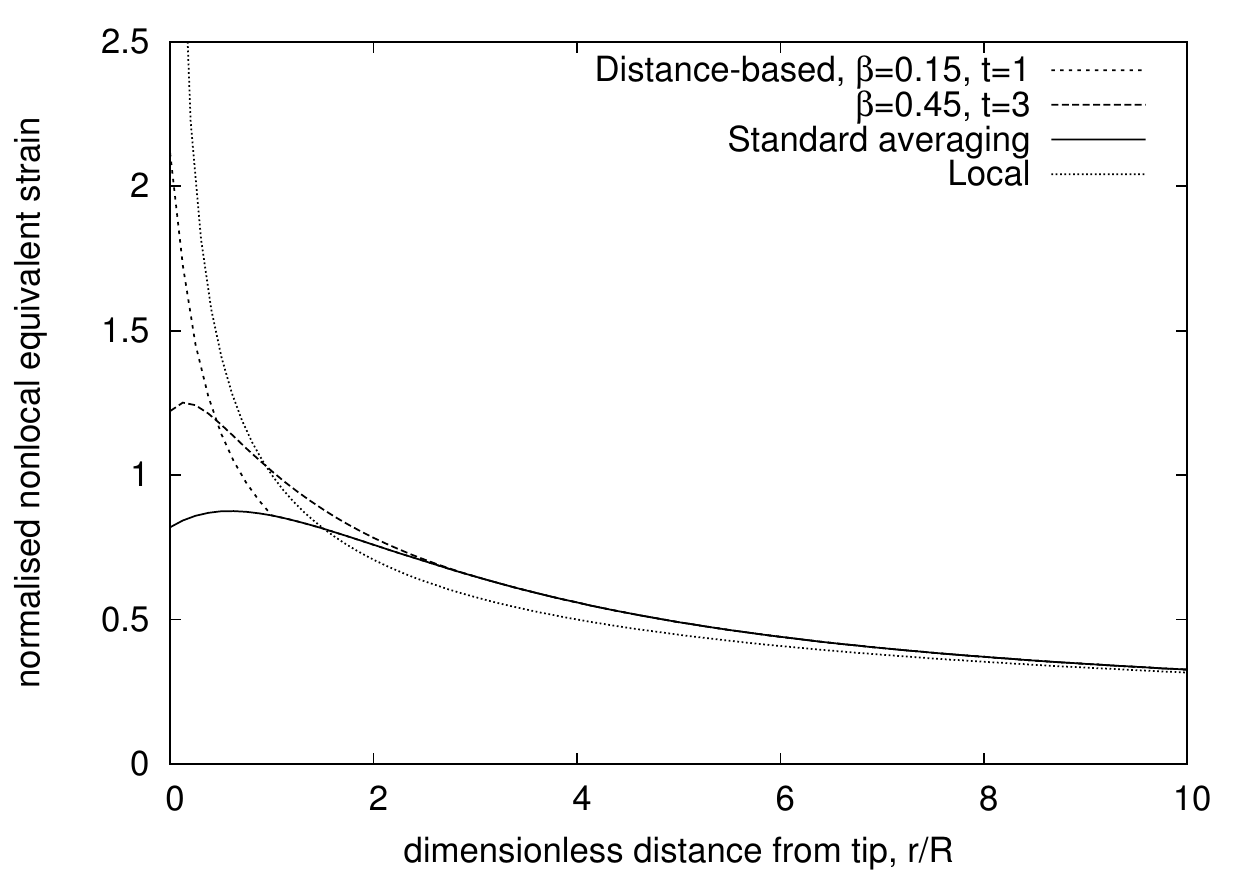}
\end{center}
\caption{Distribution of local and nonlocal equivalent strain ahead of the crack tip (computed from the asymptotic part of mode-I strain field according to linear elastic fracture mechanics using the Rankine definition of equivalent strain and exponential nonlocal weight function).}
\label{fig:cracktip}
\end{figure}

As noted e.g.\ by \cite{SimAskSlu04}, the maximum nonlocal equivalent strain computed by the standard averaging procedure in the linear elastic range is usually attained at a certain distance ahead of the notch tip, rather than directly at the tip. 
As a consequence, damage growth does not initiate from the tip but from the point at which the nonlocal equivalent strain attains its maximum. The distribution of the nonlocal Rankine-type equivalent strain along the line emanating from the crack tip and aligned with the crack is plotted in Fig.~\ref{fig:cracktip}. It has been computed from the asymptotic part of the strain field characterising a mode-I crack, using the two-dimensional exponential weight function (\ref{eq:basicAlpha0}) scaled according to (\ref{eq:basicWeight}).
The results are presented in terms of normalised quantities, with the distance from the crack tip normalised by the characteristic length $R$ and the nonlocal equivalent strain normalised by the dimensionless factor $K_I/(E\sqrt{2\pi R})$ where $K_I$ is the mode-I stress intensity factor and $E$ is Young's modulus.  
The solid curve corresponds to the standard averaging scheme and the maximum is found at distance $0.594R$ ahead of the tip.
This is similar to the result reported by \cite{SimAskSlu04}, where the nonlocal  equivalent strain was plotted for
a Mises-type definition of equivalent strain, using the Gaussian weight function. The exact distance of the maximum from the tip
was not explicitly mentioned in that reference but our calculations indicate that it would be about $0.545R$. 
Interestingly, the modified distance-based averaging scheme can shift the point
of maximum nonlocal equivalent strain and, for some parameter combinations, move it to the tip, 
as documented in Fig.~\ref{fig:cracktip} by the dashed curve
with $\beta=0.45$ and $t=3$ and dotted curve with $\beta=0.15$ and $t=1$. At distances exceeding $tR$, the nonlocal average is
the same as for the standard scheme, but at shorter distances from the tip the nonlocal value is increased, because
averaging is performed with a reduced interaction distance and the relative weight of the near neighbours with high local strain values
is augmented. Parameter $t$ sets the maximum distance $tR$ to which this effect is felt, while parameter $\beta$ controls the
nonlocal value at the tip, which scales with $1/\sqrt{\beta}$. For comparison, Fig.~\ref{fig:cracktip}
also shows the distribution of the local equivalent strain, which has a singularity at the tip; see the dash-dotted curve.

Parameter $\beta$ in both distance-based and stress-based approaches influences the amount of nonlocal interactions in the weight function $\alpha(\boldsymbol{x},\boldsymbol{\xi})$. A small value of $\beta$  reduces the interaction distance.
If $\beta$ is chosen too small, it results in a ``local'' definition of the nonlocal equivalent strain $\bar{\varepsilon}_{\rm eq}$. Consequently, extremely fine meshes are needed, otherwise damage may localise in a single band of elements resulting in irregular strain profiles and mesh-sensitive results.  
Therefore, a minimum value of $\beta = 0.15$ was enforced for the distance and stress-based approaches so that the width of the localisation zone was larger than the finite elements.
If a very fine mesh was used, $\beta$ could be further reduced and it can be expected that the stress-based approach would yield an even better agreement with the meso-scale results

\begin{figure}
\begin{center}
\begin{tabular}{cc}
\includegraphics[width=9cm]{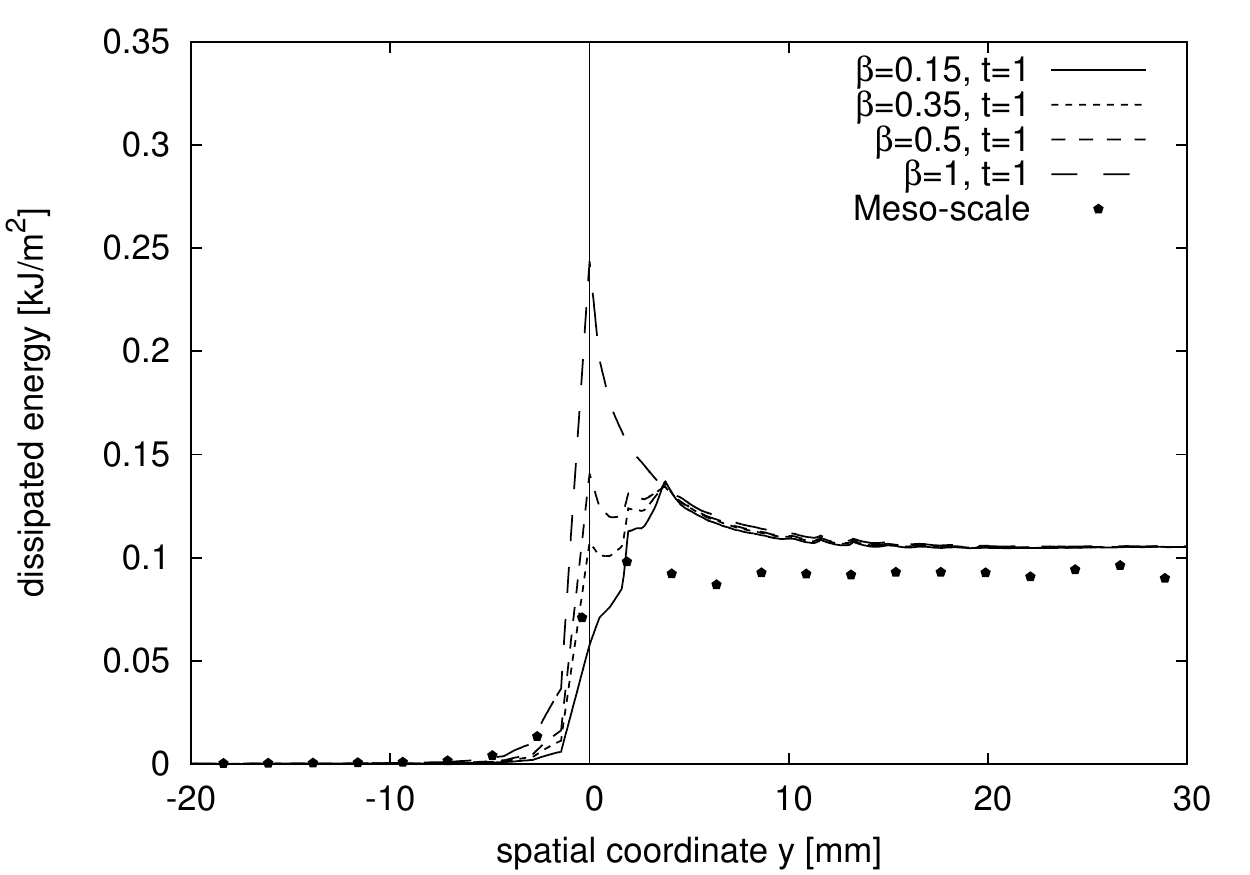} & \includegraphics[width=9cm]{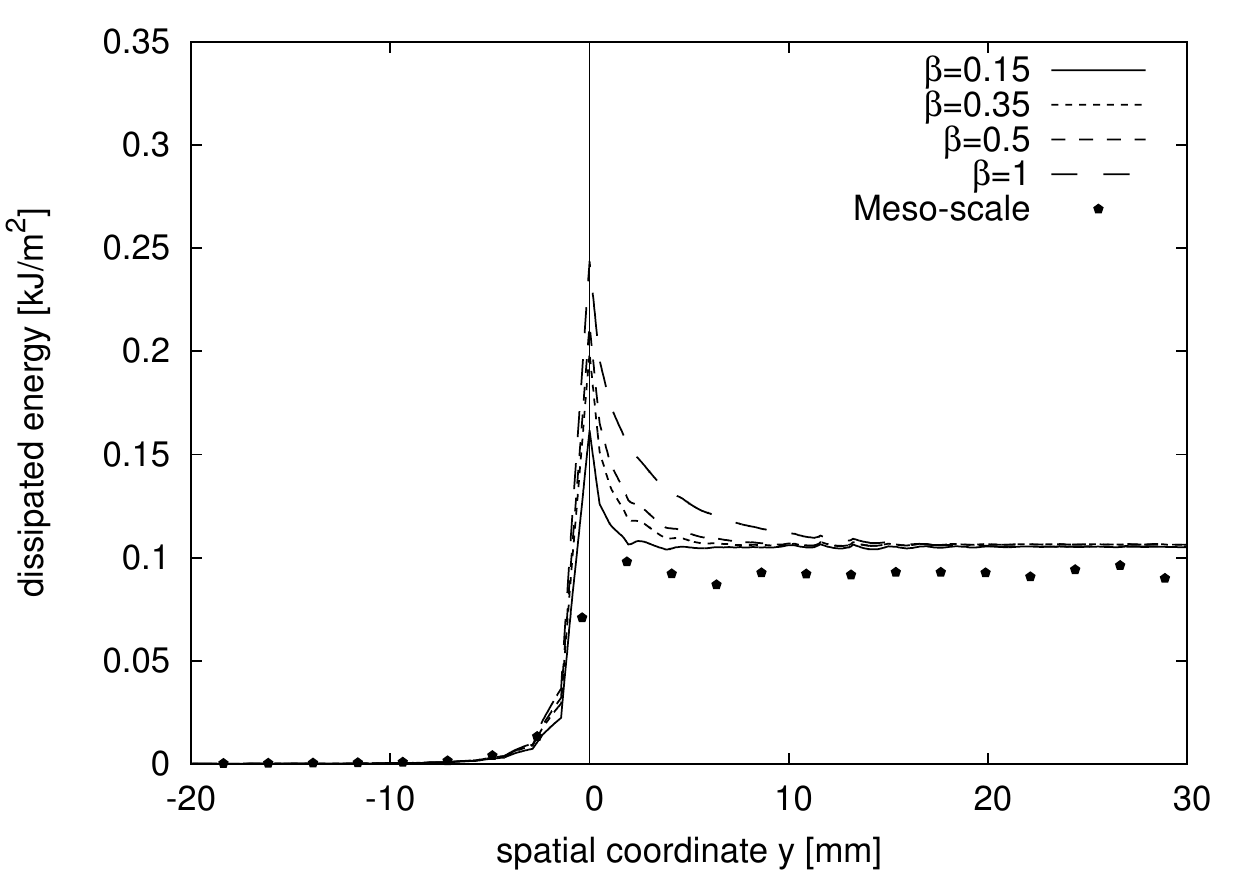}\\
(a) & (b)\\
\end{tabular}
\end{center}
\caption{Effect of parameter $\beta$ on the dissipated energy profiles along the ligament length for the V-notched specimen ($\alpha = 45^{\circ}$) and the
(a) distance-based approach, (b) stress-based damage approach.}
\label{fig:2dBeta}
\end{figure}

For illustrating the influence of parameter $\beta$ on the dissipation for the stress-based and distance-based models, the dissipated energy distribution along the ligament length for the V-notch is shown in Fig.~\ref{fig:2dBeta}. Again, the specimen with the V-notch is used as it represents an intermediate case between sharply notched and unnotched specimens.
For both models, the parameter $\beta$ has a strong influence on the dissipation above the notch. 
The smaller $\beta$, the stronger is the reduction of the dissipated energy. 
However, the influence differs for these two models.
For the stress-based model, the shape of the dissipation distribution is almost independent of $\beta$ (Fig.~\ref{fig:2dBeta}a) and
the peak value of dissipated energy density is reduced. 
For the distance-based model, the peak value of the dissipated energy is also reduced, but the parameter $\beta$ has no influence beyond the distance $t R = 4$~mm (Fig.~\ref{fig:2dBeta}b).

\begin{figure}
\begin{center}
\begin{tabular}{cc}
\includegraphics[width=9cm]{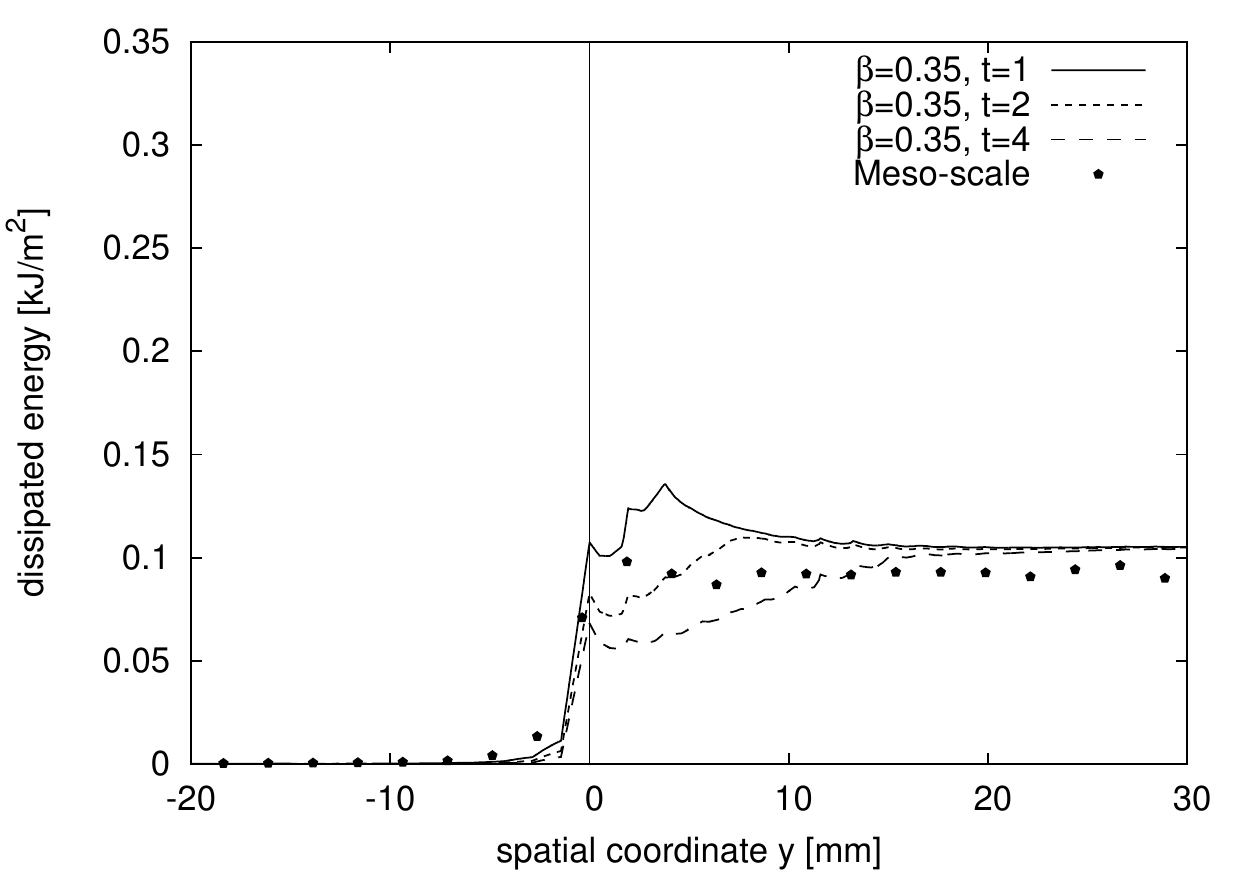}&
\includegraphics[width=9cm]{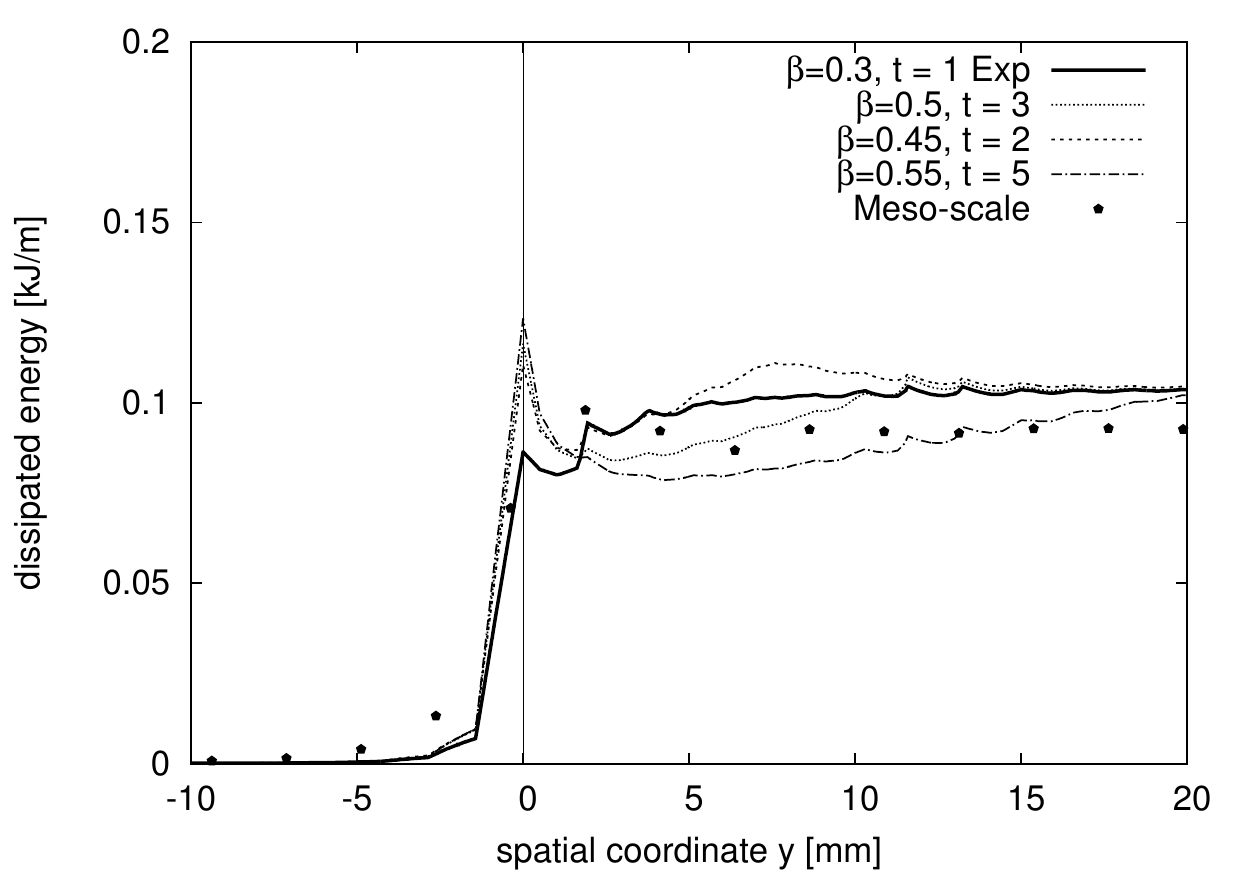}\\
(a) & (b)\\
\end{tabular}
\end{center}
\caption{Effect of parameters of the distance-based approach on the dissipated energy profiles along the ligament length for the V-notched specimen ($\alpha = 45^{\circ}$): (a) variation of parameter $t$ at fixed $\beta=0.35$, (b) combined variation of parameters $t$ and $\beta$.}
\label{fig:2dT}
\end{figure}

The first conclusion of this parameter study is that, for the stress-based approach, parameter $\beta$ should be chosen as small as possible, same as in \cite{GirDufMaz11}. 
On the other hand, for the distance-based approach, a small $\beta$ in combination with a large $tR$ may result in an underestimation of the dissipated energy.
The influence of parameter $t$ is documented in Fig.~\ref{fig:2dT}a, which shows
the distribution of dissipated energy along the ligament of the V-notched specimen for $t=1$, 2 and 4, with fixed $\beta=0.35$ and $R=4$ mm. As expected, the reduction of dissipation near the notch is stronger for higher values of $t$, and the zone affected by the reduction extends up to distance $tR$ from the notch tip. 
To compensate for the effect of increased parameter $t$ on dissipation density near the notch tip, it seems reasonable to use larger values of $\beta$ than in Fig.~\ref{fig:2dT}a.
However, as illustrated by the dashed and dotted curves in Fig.~\ref{fig:2dT}b, values of $\beta>0.35$ typically lead to the formation of a local dissipation peak at the notch tip, even for values of $t$ larger than 1. With properly chosen combinations of parameters $\beta$ and $t$, the dissipation near the notch is correct on the average but its distribution is still somewhat irregular. A further improvement is achieved by using a smooth dependence of the reduction factor $\gamma$ on the distance from the boundary, instead of the piecewise linear dependence according to formula (\ref{eq:gamma}).
The exponential formula
\begin{equation}\label{eq:gammamod}
\gamma(\boldsymbol{x}) = 1 - (1-\beta)\exp\left(-\frac{d(\boldsymbol{x})}{tR}\right) 
\end{equation}
still uses just two parameters, same as the piecewise linear formula (\ref{eq:gamma}), but it can eliminate the local peak and provide a very good overall shape of the dissipation density distribution; see the thick solid curve in  Fig.~\ref{fig:2dT}b, obtained with parameters $\beta=0.3$ and $t=1$. Note that, for the exponential formula (\ref{eq:gammamod}), the effect of the boundary on the reduction of the nonlocal interaction distance does not vanish for $d(\boldsymbol{x})\ge tR$, and so the value $t=1$, which was too small for the  piecewise linear formula (\ref{eq:gamma}), turns out to be appropriate.

All the results presented here have been obtained for a specific isotropic damage model, with a Rankine-like expression for equivalent strain.
A different choice of the equivalent strain expression might lead to higher or lower dissipation under multiaxial stress but would not affect the shape of the dissipation distribution along the ligament. In other words, the conclusions regarding the excessive dissipation near the notch are valid for other choices, too.

\subsubsection{Nonlocal damage-plastic models}

\begin{figure}
\begin{center}
\begin{tabular}{cc}
\includegraphics[width=9cm]{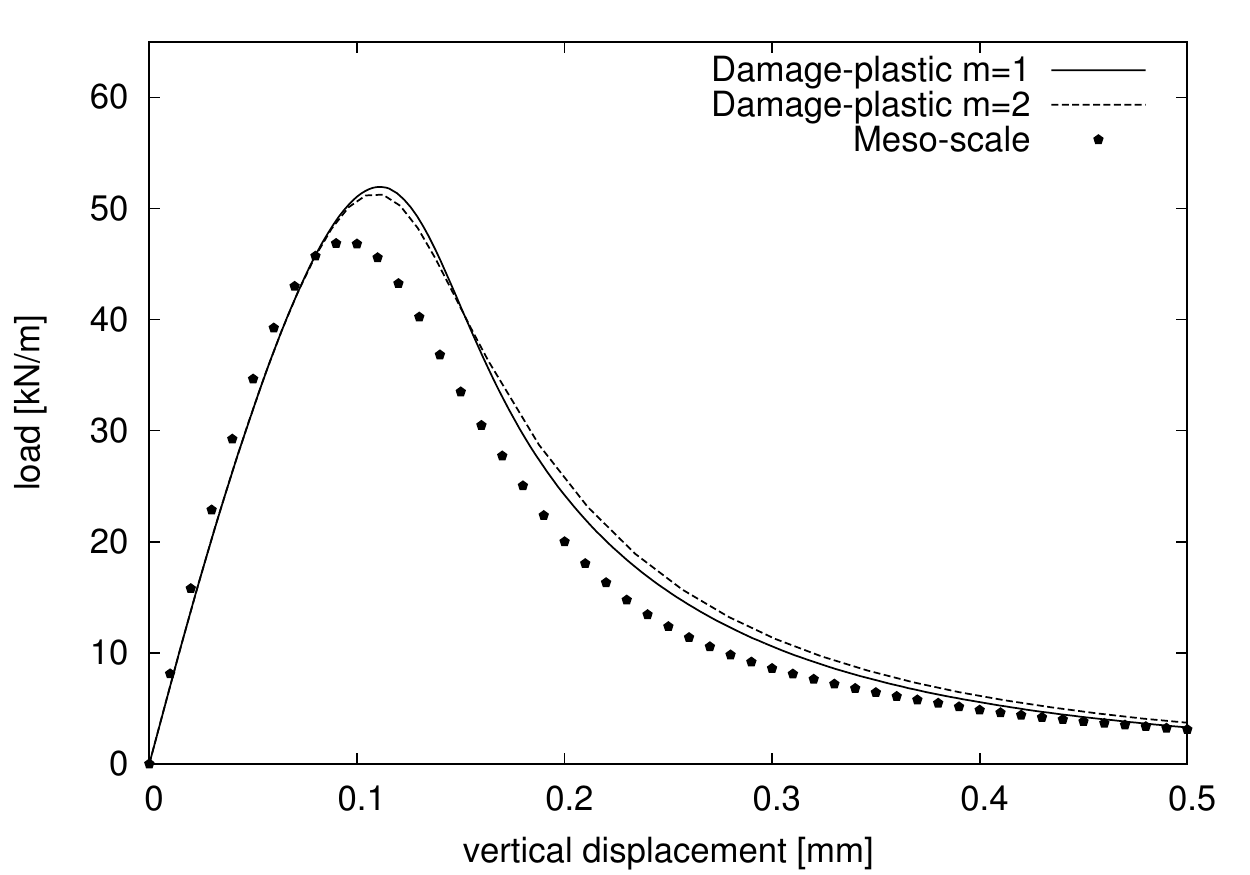}  & \includegraphics[width=9cm]{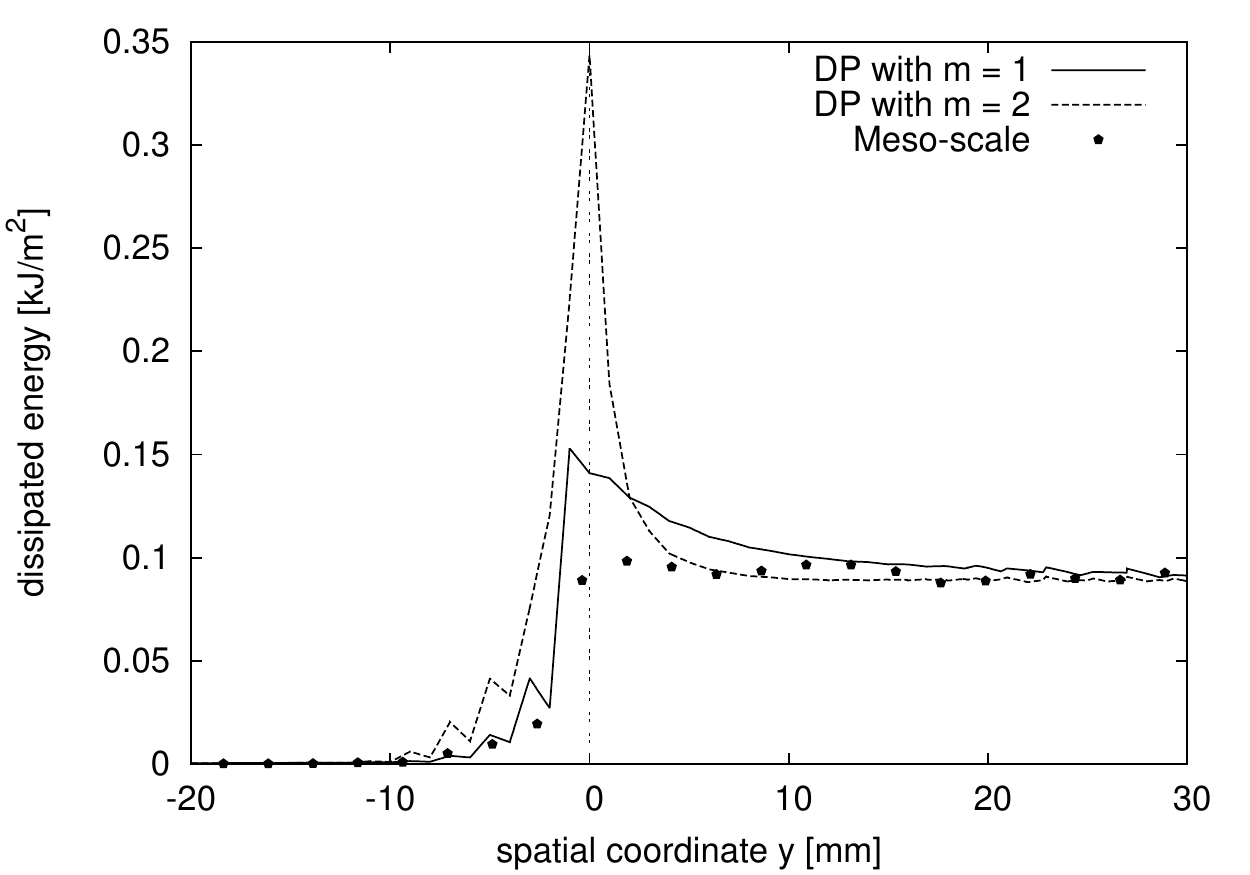}\\
(a) & (b)
\end{tabular}
\end{center}
\caption{Comparison of the results of the {\bf nonlocal damage-plastic} approach with $m=1$~and~$2$, and meso-scale analysis for specimen with a {\bf sharp notch} ($\alpha = 0^{\circ}$): (a) load-displacement curves and (b) dissipated energy profiles.}
\label{fig:2d0b}
\end{figure}

\begin{figure}
\begin{center}
\begin{tabular}{cc}
\includegraphics[width=9cm]{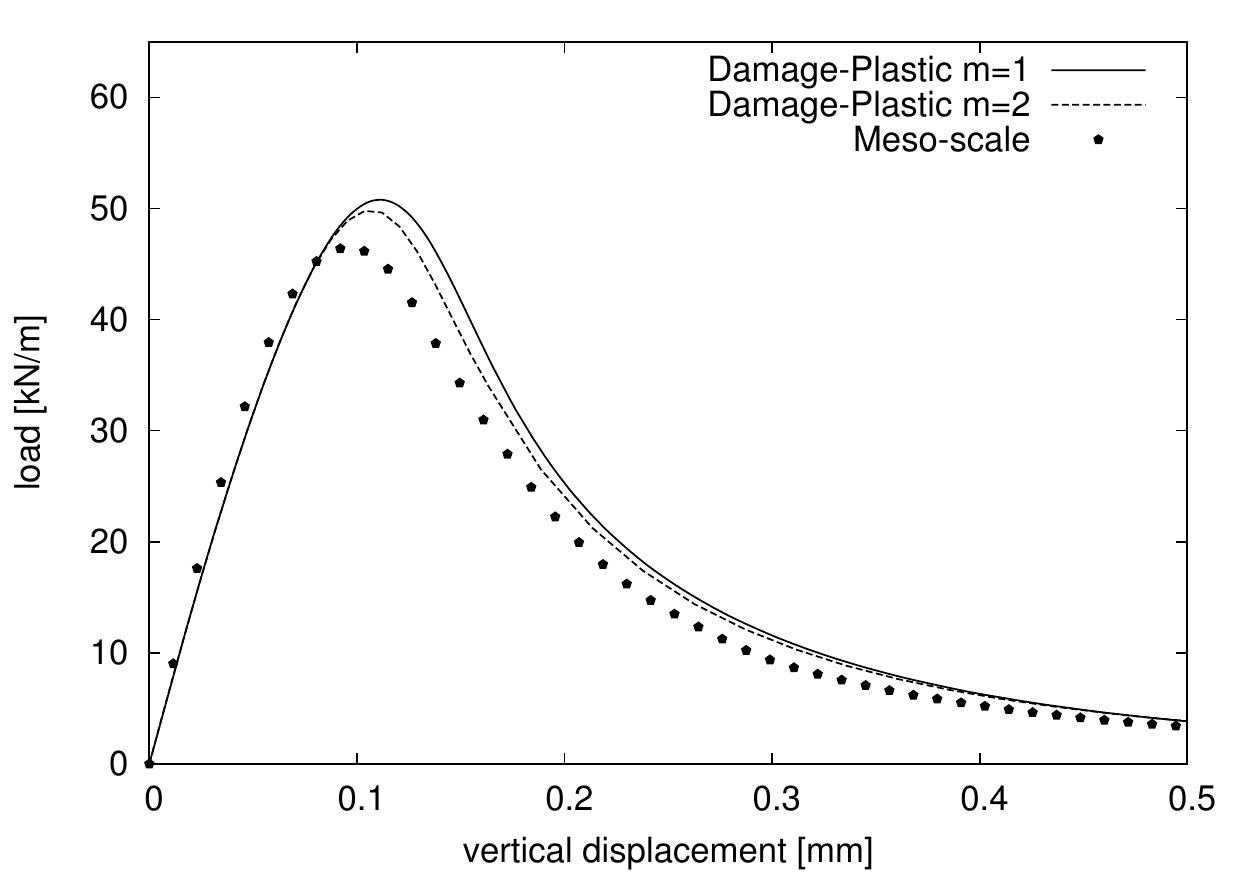} & \includegraphics[width=9cm]{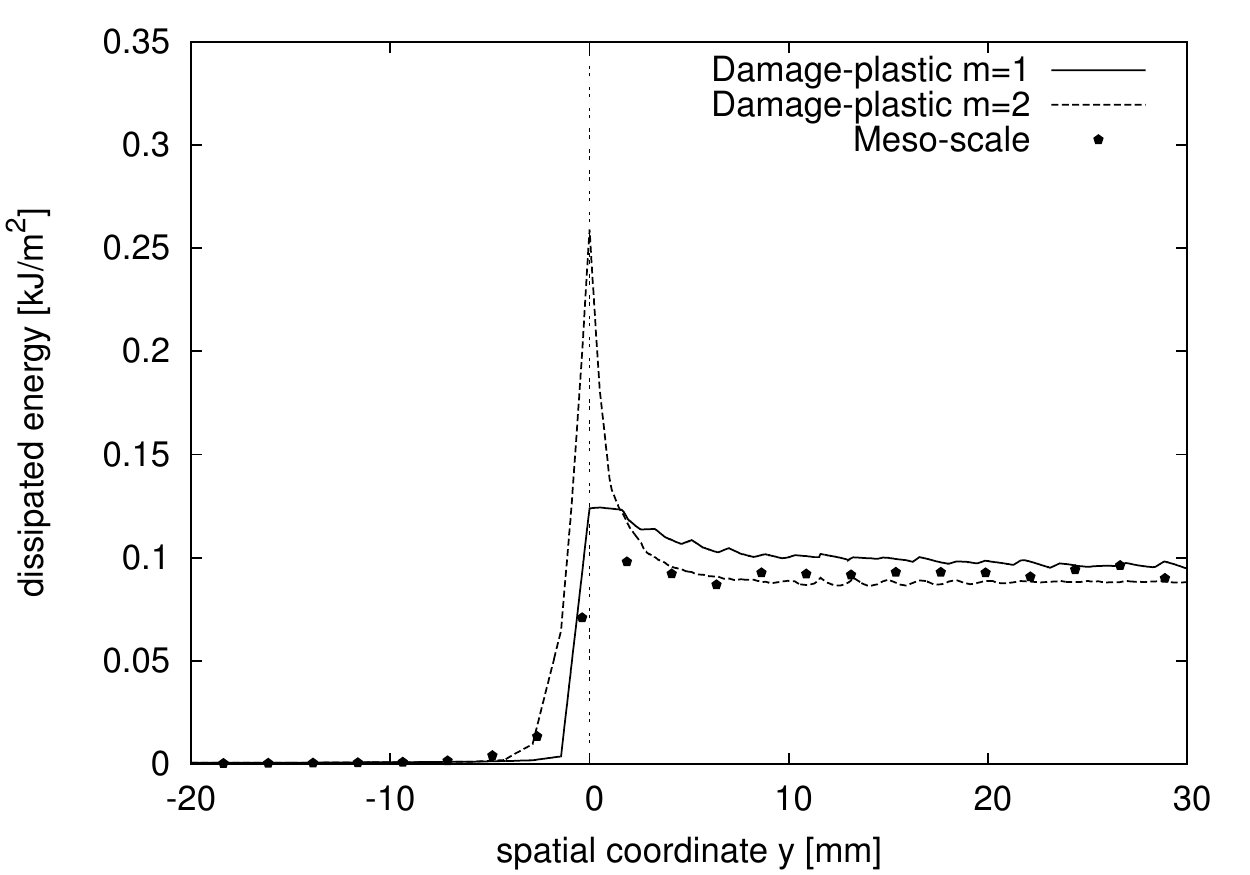}\\
(a) & (b)\\
\end{tabular}
\end{center}
\caption{Comparison of the results of the {\bf nonlocal damage-plastic} approach with $m=1$~and~$2$ and meso-scale analysis for specimen with a {\bf V-notch} ($\alpha = 45^{\circ}$): (a) load-displacement curves and (b) dissipated energy profiles.}
\label{fig:2d45b}
\end{figure}

\begin{figure}
\begin{center}
\begin{tabular}{cc}
\includegraphics[width=9cm]{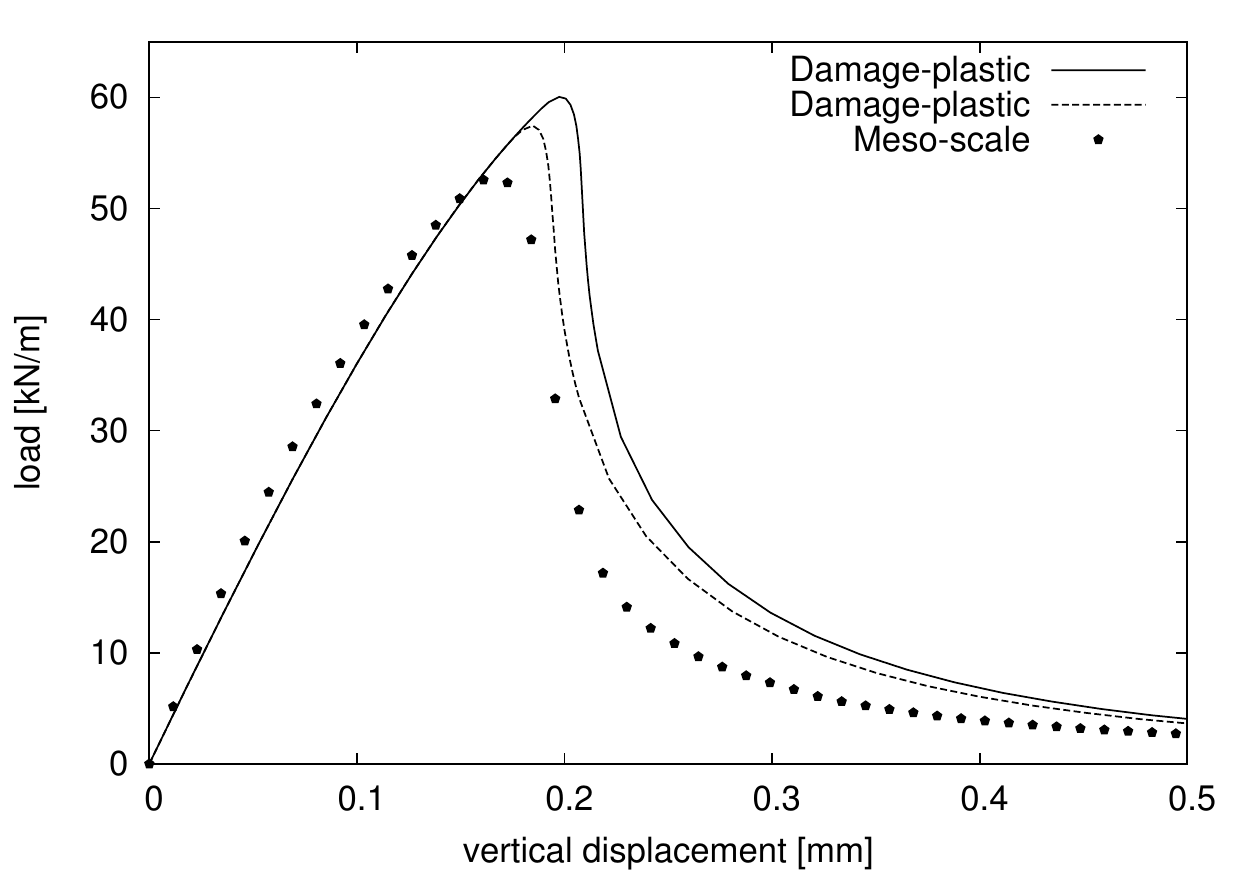} & \includegraphics[width=9cm]{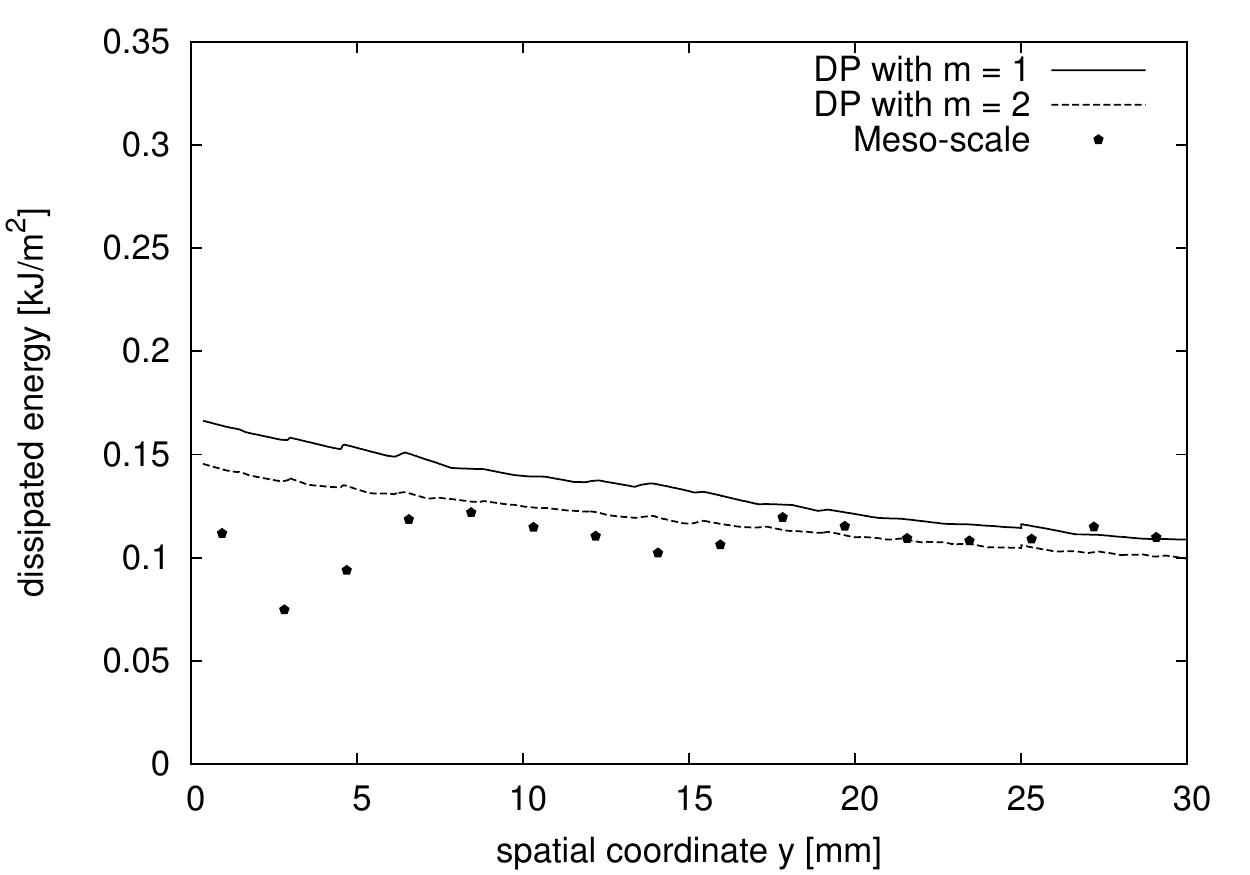}\\
(a) & (b)\\
\end{tabular}
\end{center}
\caption{Comparison of the results of the {\bf nonlocal damage-plastic} approach with $m=1$~and~$2$ and meso-scale analysis for the {\bf unnotched} specimen ($\alpha = 90^{\circ}$): (a) load-displacement curves and (b) dissipated energy profiles.}
\label{fig:2d90b}
\end{figure}

The results obtained with the nonlocal damage-plastic approaches for the three beam geometries are compared to meso-scale results in Figs.~\ref{fig:2d0b}--\ref{fig:2d90b} in the form of load-displacement curves and dissipated energy distribution along the ligament of the beam. 
For the damage-plastic model, the maximum principal stress cannot exceed the tensile strength, 
so one could expect that the dissipation near the notch is lower than for the nonlocal damage models.
However, it turns out that the over-nonlocal damage-plastic model with $m=2$ exhibits a high
peak of dissipation density at the notch, comparable to the nonlocal damage model with
standard averaging. For the nonlocal damage-plastic model with $m=1$, dissipation near
the notch is only slightly above the value in the regular part of the ligament, far from the notch.
 
\begin{figure}
\begin{center}
\begin{tabular}{ccc}
 \includegraphics[width=9cm]{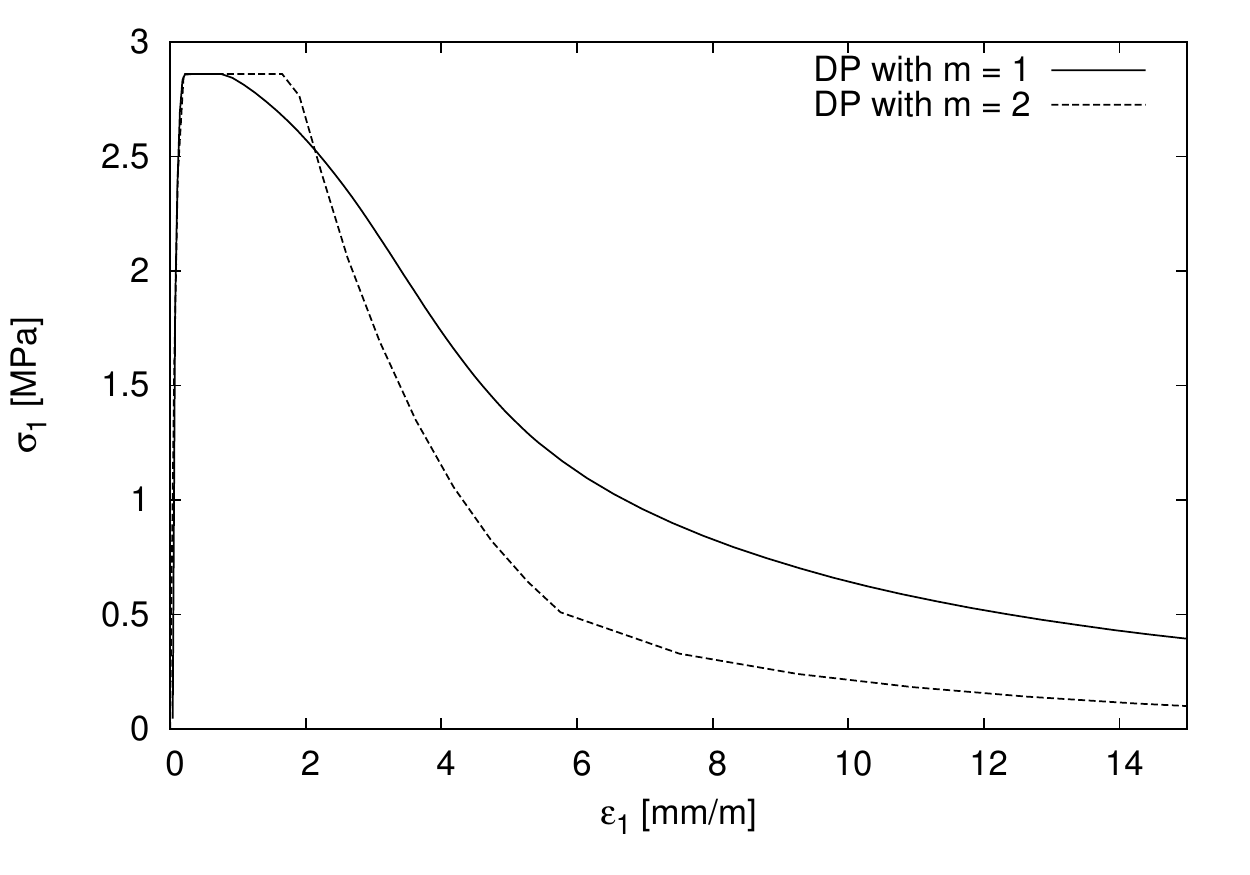} & \includegraphics[width=3cm]{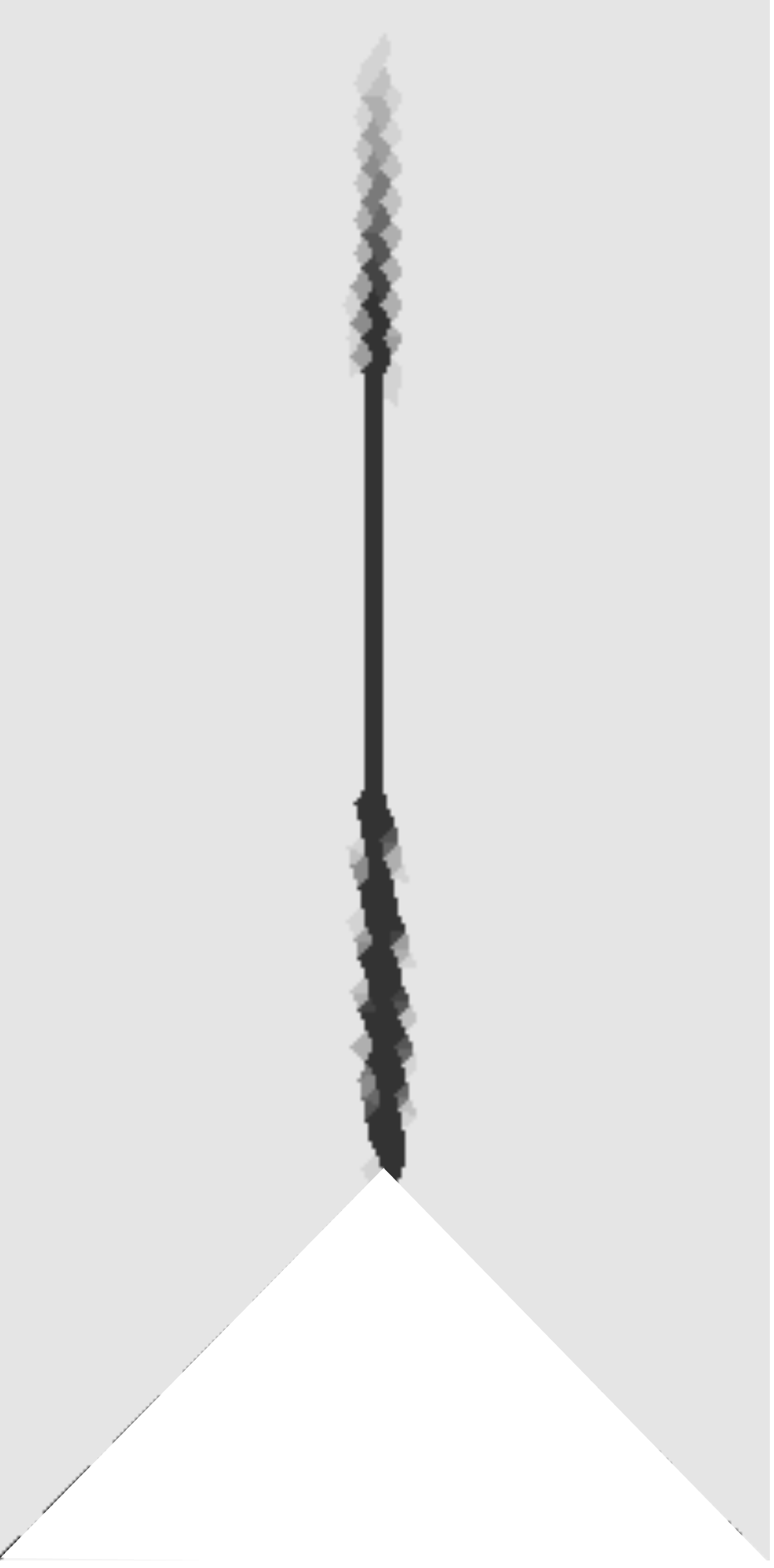} & \includegraphics[width=3cm]{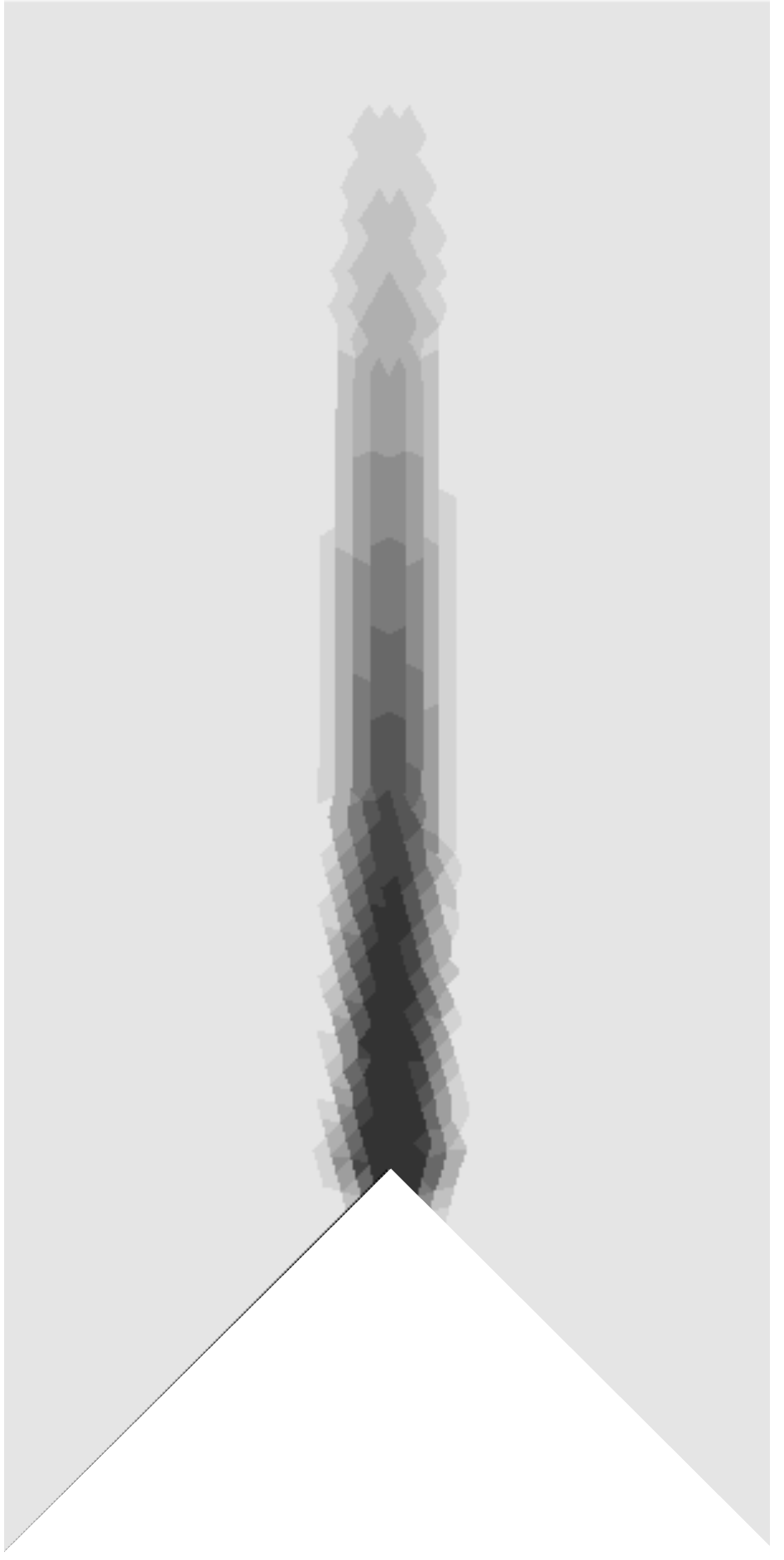}\\
(a) & (b) & (c)\\
\end{tabular}
\end{center}
\caption{Results for the damage-plastic model for the V-notched specimen ($\alpha = 45^{\circ}$) in the form of (a) maximum principal stress versus strain for an element just above the notch, (b)--(c) contour plots of the local plastic hardening parameter for  (b) $m=1$ and (c) $m=2$  at a displacement of 1.88~mm in Fig.~\ref{fig:2d45b}. Black indicates strains greater than 0.1. Only a part of the depth of the beam is shown.}
\label{fig:contourdp}
\end{figure}

Since the effective stress is computed using an elasto-plastic law with limited hardening, it cannot exceed the specified tensile strength.
This is demonstrated for one element just above the notch in Fig.~\ref{fig:contourdp}a by a plot of the maximum principal stress versus maximum principal strain.
Nonlocality does not lead to excessive stresses but still, the dissipation density increases compared to the corresponding local model because the damage is driven by the
nonlocal plastic strain, which lags behind the local plastic strain.
The over-nonlocal damage-plastic model ($m=2$) enforces nonlocality at the notch tip, which ensures that a distributed (and mesh independent) plastic strain profile is obtained (Fig.~\ref{fig:contourdp}c).
The standard nonlocal damage-plastic model ($m=1$) produces energy dissipation in much agreement with the meso-scale results (Fig.~\ref{fig:2d45b}b). However, the strain profile is localised (Fig.~\ref{fig:contourdp})b.


\section{Conclusions}

In the present study, a nonlocal isotropic damage model with different averaging procedures and a nonlocal damage-plastic model were applied to the modelling of fracture in three-point bending tests with different notch geometries.
In the analyses of the sharp notched and V-notched beams, the {\bf damage} model with the standard scaling approach overpredicts the energy dissipation in the vicinity of the notch, which results in an overestimation of the load-carrying capacity. For the damage model with local complement and with the stress-based and distance-based averaging approaches, the energy dissipation close to the notch is reduced, which gives a better agreement with the meso-scale results. 
For the unnotched case, the dissipated energy is distributed in a reasonable way for all approaches, but is somewhat too low for the averaging method based on local complement.
The distance-based approach requires two additional input parameters (compared to the standard scaling), whereas the stress-based approach requires only one additional parameter and the modification based on local complement does not require any additional parameter.

The nonlocal {\bf damage-plastic} approach does not require any additional parameters, but leads to an overestimation of the dissipated energy close to the notch, if the over-nonlocal averaging approach is used. For a standard averaging with $m=1$, the damage-plasticity model gives results which are in better agreement with the meso-scale analyses. However, for this damage-plastic model, the width of the fracture process zone depends on the size of the elements used.

An alternative regularisation technique that deserves attention is the implicit gradient formulation \citep{Peerlings96}, which defines the nonlocal field as the solution of a boundary value problem. Its effect on energy dissipation near nonconvex boundaries will be addressed in future work.

\section*{Acknowledgement}

The first and second authors acknowledge funding received from the UK’s Engineering and Physical Sciences Research Council (EPSRC) under grant EP/I036427/1 and EP/J500434/1, respectively.
The third and fourth authors gratefully acknowledge financial support received from the Czech Science Foundation (GA\v{C}R) under project 108/11/1243.
\bibliographystyle{plainnat}

\bibliography{general}

\begin{thebibliography}{32}
\providecommand{\natexlab}[1]{#1}
\providecommand{\url}[1]{\texttt{#1}}
\expandafter\ifx\csname urlstyle\endcsname\relax
  \providecommand{\doi}[1]{doi: #1}\else
  \providecommand{\doi}{doi: \begingroup \urlstyle{rm}\Url}\fi

\bibitem[Ba\v{z}ant(1994)]{Baz94}
Z.~P. Ba\v{z}ant.
\newblock Nonlocal damage theory based on micromechanics of crack interactions.
\newblock \emph{\JEM}, 120:\penalty0 593--617, 1994.

\bibitem[Ba\v{z}ant and Jir\'{a}sek(2002)]{BazJir02}
Z.~P. Ba\v{z}ant and M.~Jir\'{a}sek.
\newblock Nonlocal integral formulations of plasticity and damage: {S}urvey of
  progress.
\newblock \emph{\JEM}, 128\penalty0 (10):\penalty0 1119--1149, 2002.

\bibitem[Ba\v{z}ant et~al.(2010)Ba\v{z}ant, Le, and Hoover]{BazLeHo10}
Z.~P. Ba\v{z}ant, J.-L. Le, and C.~G. Hoover.
\newblock Nonlocal boundary layer (nbl) model:overcoming boundary condition
  problems in strength statistics and fracture analysis of quasibrittle
  materials.
\newblock In \emph{Fracture Mechanics of Concrete and Concrete Structures},
  pages 135--143, JeJu, Korea, 2010.

\bibitem[Bolander and Hikosaka(1995)]{BolHik95}
J.~Bolander and H.~Hikosaka.
\newblock Simulation of fracture in cement-based composites.
\newblock \emph{Cement and Concrete Composites}, 17:\penalty0 135--145, 1995.

\bibitem[Bolander and Saito(1998)]{BolSai98}
J.~E. Bolander and S.~Saito.
\newblock Fracture analysis using spring networks with random geometry.
\newblock \emph{Engineering Fracture Mechanics}, 61:\penalty0 569--591, 1998.

\bibitem[Borino et~al.(2002)Borino, Failla, and Parrinello]{BorFaiPar02}
G.~Borino, B.~Failla, and F.~Parrinello.
\newblock A symmetric formulation for nonlocal damage models.
\newblock In H.~A. Mang, F.~G. Rammerstorfer, and J.~Eberhardsteiner, editors,
  \emph{Proceedings of the Fifth World Congress on Computational Mechanics
  (WCCM V)}, Vienna, Austria, 2002. Vienna University of Technology.
\newblock ISBN 3-9501554-0-6, http://wccm.tuwien.ac.at.

\bibitem[Borino et~al.(2003)Borino, Failla, and Parrinello]{BorFaiPar03}
G.~Borino, B.~Failla, and F.~Parrinello.
\newblock {A symmetric nonlocal damage theory}.
\newblock \emph{International Journal of Solids and Structures}, 40\penalty0
  (13-14):\penalty0 3621--3645, 2003.

\bibitem[Cundall and Strack(1979)]{Cun79}
P.~A. Cundall and O.~D.~L. Strack.
\newblock A discrete numerical model for granular assemblies.
\newblock \emph{G\'{e}otechnique}, 29:\penalty0 47--65, 1979.

\bibitem[Giry et~al.(2011)Giry, Dufour, and Mazars]{GirDufMaz11}
C.~Giry, F.~Dufour, and J.~Mazars.
\newblock Stress-based nonlocal damage model.
\newblock \emph{International Journal of Solids and Structures}, 48:\penalty0
  3431--3443, 2011.

\bibitem[Grassl(2009)]{Gra09b}
P.~Grassl.
\newblock On a damage-plasticity approach to model concrete failure.
\newblock \emph{Proceedings of the ICE - Engineering and Computational
  Mechanics}, 162:\penalty0 221--231, 2009.

\bibitem[Grassl and Jir{\'a}sek(2006)]{GraJir06}
P.~Grassl and M.~Jir{\'a}sek.
\newblock {Damage-plastic model for concrete failure}.
\newblock \emph{International Journal of Solids and Structures}, 43:\penalty0
  7166--7196, 2006.

\bibitem[Grassl and Jir\'{a}sek(2006)]{GraJir06a}
P.~Grassl and M.~Jir\'{a}sek.
\newblock A plastic model with nonlocal damage applied to concrete.
\newblock \emph{International Journal for Numerical and Analytical Methods in
  Geomechanics}, 30:\penalty0 71--90, 2006.

\bibitem[Grassl and Jir\'{a}sek(2010)]{GraJir10}
P.~Grassl and M.~Jir\'{a}sek.
\newblock Meso-scale approach to modelling the fracture process zone of
  concrete subjected to uniaxial tension.
\newblock \emph{International Journal of Solids and Structures}, 47:\penalty0
  957--968, 2010.

\bibitem[Grassl et~al.(2012)Grassl, Gr\'{e}goire, Solano, and
  Pijaudier-Cabot]{GraGreSol12}
P.~Grassl, D.~Gr\'{e}goire, L.~R. Solano, and G.~Pijaudier-Cabot.
\newblock Meso-scale modelling of the size effect on the fracture process zone
  of concrete.
\newblock \emph{International Journal of Solids and Structures}, 49\penalty0
  (13):\penalty0 1818--1827, 2012.

\bibitem[Herrmann et~al.(1989)Herrmann, Hansen, and Roux]{HerHanRou89}
H.J. Herrmann, A.~Hansen, and S.~Roux.
\newblock {Fracture of disordered, elastic lattices in two dimensions}.
\newblock \emph{Physical Review B}, 39\penalty0 (1):\penalty0 637--648, 1989.

\bibitem[Hsu and Slate(1963)]{HsuSla63}
T.~T.~C. Hsu and F.~O. Slate.
\newblock {Tensile Bond Strength Between Aggregate and Cement Paste or Mortar}.
\newblock \emph{ACI Journal Proceedings}, 60\penalty0 (4), 1963.

\bibitem[Jir\'asek and Bauer(2012)]{JirBau12}
M.~Jir\'asek and M.~Bauer.
\newblock Numerical aspects of the crack band approach.
\newblock \emph{Computers and Structures}, 110--111:\penalty0 60--78, 2012.

\bibitem[Jir\'{a}sek and Ba\v{z}ant(1994)]{JirBaz94}
M.~Jir\'{a}sek and Z.~P. Ba\v{z}ant.
\newblock Localization analysis of nonlocal model based on crack interactions.
\newblock \emph{\JEM}, 120:\penalty0 1521--1542, 1994.

\bibitem[Jir\'{a}sek and Rolshoven(2003)]{JirRol02}
M.~Jir\'{a}sek and S.~Rolshoven.
\newblock Comparison of integral-type nonlocal plasticity models for
  strain-softening materials.
\newblock \emph{\IJES}, 41:\penalty0 1553--1602, 2003.

\bibitem[Jir\'{a}sek et~al.(2004)Jir\'{a}sek, Rolshoven, and
  Grassl]{JirRolGra04}
M.~Jir\'{a}sek, S.~Rolshoven, and P.~Grassl.
\newblock Size effect on fracture energy induced by non-locality.
\newblock \emph{International Journal of Numerical and Analytical Methods in
  Geomechanics}, 28:\penalty0 653--670, 2004.

\bibitem[Kawai(1978)]{Kaw78}
T.~Kawai.
\newblock {New discrete models and their application to seismic response
  analysis of structures}.
\newblock \emph{Nuclear Engineering and Design}, 48:\penalty0 207--229, 1978.

\bibitem[Krayani et~al.(2009)Krayani, Pijaudier-Cabot, and Dufour]{KraPijDuf09}
A.~Krayani, G.~Pijaudier-Cabot, and F.~Dufour.
\newblock Boundary effect on weight function in nonlocal damage model.
\newblock \emph{Engineering Fracture Mechanics}, pages 2217--2231, 2009.

\bibitem[Patz\'ak(2012)]{Pat12}
B.~Patz\'ak.
\newblock {OOFEM} -- an object-oriented simulation tool for advanced modeling
  of materials and structures.
\newblock \emph{Acta Polytechnica}, 52:\penalty0 59--66, 2012.

\bibitem[Patz\'{a}k and Bittnar(2001)]{PatBit01}
B.~Patz\'{a}k and Z.~Bittnar.
\newblock Design of object oriented finite element code.
\newblock \emph{Advances in Engineering Software}, 32:\penalty0 759--767, 2001.

\bibitem[Peerlings et~al.(1996)Peerlings, de~Borst, Brekelmans, and
  de~Vree]{Peerlings96}
R.~H.~J. Peerlings, R.~de~Borst, W.~A.~M. Brekelmans, and J.~H.~P. de~Vree.
\newblock Gradient-enhanced damage for quasi-brittle materials.
\newblock \emph{\IJNME}, 39:\penalty0 3391--3403, 1996.

\bibitem[Pijaudier-Cabot and Ba\v{z}ant(1987)]{PijBaz87}
G.~Pijaudier-Cabot and Z.~P. Ba\v{z}ant.
\newblock Nonlocal damage theory.
\newblock \emph{\JEM}, 113:\penalty0 1512--1533, 1987.

\bibitem[Pijaudier-Cabot and Dufour(2010)]{PijDuf10}
G.~Pijaudier-Cabot and F.~Dufour.
\newblock Non local damage model. {B}oundary and evolving boundary effects.
\newblock \emph{European Journal of Environmental and Civil Engineering},
  14:\penalty0 729--749, 2010.

\bibitem[Polizzotto(2002)]{Pol02}
C.~Polizzotto.
\newblock Remarks on some aspects of nonlocal theories in solid mechanics.
\newblock In \emph{Proceedings of the 6th National Congress SIMAI}, Chia
  Laguna, Italy, May 2002.
\newblock CD-ROM.

\bibitem[Schlangen and van Mier(1992)]{SchMie92b}
E.~Schlangen and J.~G.~M. van Mier.
\newblock Simple lattice model for numerical simulation of fracture of concrete
  materials and structures.
\newblock \emph{Materials and Structures}, 25:\penalty0 534--542, 1992.

\bibitem[Simone et~al.(2004)Simone, Askes, and Sluys]{SimAskSlu04}
A.~Simone, H.~Askes, and L.~J. Sluys.
\newblock Incorrect initiation and propagation of failure in non-local and
  gradient-enhanced media.
\newblock \emph{International Journal of Solids and Structures}, 41\penalty0
  (2):\penalty0 351--363, 2004.

\bibitem[Str\"{o}mberg and Ristinmaa(1996)]{Stromberg96}
L.~Str\"{o}mberg and M.~Ristinmaa.
\newblock {FE}-formulation of a nonlocal plasticity theory.
\newblock \emph{\CMAME}, 136:\penalty0 127--144, 1996.

\bibitem[Vermeer and Brinkgreve(1994)]{VerBri94}
P.~A. Vermeer and R.~B.~J. Brinkgreve.
\newblock A new effective non-local strain measure for softening plasticity.
\newblock In R.~Chambon, J.~Desrues, and I.~Vardoulakis, editors,
  \emph{Localisation and Bifurcation Theory for Soils and Rocks}, pages
  89--100, Rotterdam, 1994. Balkema.

\end{thebibliography}

\end{document}